\documentstyle[epsf,12pt]{article}
\newcommand{\yourabstract}[1]{
\mbox{}\\
\mbox{}\\
{\bf\noindent Abstract}\\
\begin{center}
\mbox{}\parbox[t]{5.in}{#1}
\end{center} }
\newcommand{\yoursection}[1]{
\vskip 2\baselineskip
{\bf\noindent #1}\\
\mbox{}\\
\vspace{-0.19in} }

\newcommand{\half}{\textstyle{\frac{1}{2}}}
\newcommand{\quarter}{\textstyle{\frac{1}{4}}}
\newcommand{\eighth}{\textstyle{\frac{1}{8}}}
\newcommand{\beq}{\begin{equation}}
\newcommand{\beqar}{\begin{eqnarray}}
\newcommand{\eeq}[1]{\label{#1} \end{equation}}
\newcommand{\eeqar}[1]{\label{#1} \end{eqnarray}}
\def\pra#1{{Phys. Rev. }{A#1} }

\def\prd#1{{Phys. Rev. }{D#1} }
\def\prl#1{{Phys. Rev. Lett. }{#1} }
\def\pl#1{{Phys. Lett. }{#1} }
\def\np#1{{Nucl. Phys. }{#1} }

\def\prep#1{{Phys. Rep. }{#1} }

\def\zp#1{{Z. Phys. }{#1} }

\def\rmp#1{{Rev. Mod. Phys. }{#1} }

\begin{document}
\begin{titlepage}
\begin{flushright}
CERN-TH.7131/93 \\
\end{flushright}
\vskip 2.0cm
\begin{center}
{\bf \large Quantum Decoherence, Entropy and Thermalization}
\vskip 0.3cm
{\bf \large in Strong Interactions at High Energy}
           \footnote{
           Work supported by the Heisenberg
           Programme (Deutsche Forschungsgemeinschaft).
           }
\vskip 0.5cm
{\bf I. Noisy and Dissipative Vacuum Effects in Toy Models}
\vskip 0.8cm
Hans-Thomas Elze
\vskip 0.3cm
CERN-Theory,
CH-1211 Geneva 23,
Switzerland \footnote{ELZE@CERNVM.CERN.CH}
\end{center}
\yourabstract{Entropy is generated in
high-multiplicity events by a dynamical separation of
strongly interacting systems into partons and unobservable
environment modes (almost constant field configurations) due to
confinement.
The effect is demonstrated in a
non-relativistic single-particle model and a scalar field theory, where
it amounts to quantum field Brownian motion. We
analyze the quantum decoherence of partons, which formally corresponds
to non-unitary time-evolution and causes entropy production, in terms
of Schmidt and pointer states in the non-relativistic case. For
the coupled scalar fields (partons and environment) we derive the
Cornwall-Jackiw-Tomboulis effective action and equations of motion
with the non-perturbative
time-dependent variational method in TDHF approximation
(to study the time-evolution of model structure functions in
the sequel). We obtain a model-independent lower bound for the entropy
in terms of two-point Wightman functions.
}
\vskip 1.4cm
\begin{center}
{\it Submitted to Nuclear Physics B.}
\end{center}
\vskip 0.9cm
\flushleft{CERN-TH.7131/93

March 1994}
\end{titlepage}

\yoursection{1. Introduction}

The idea of highly excited hadronic matter can be traced back to the
earliest observations of high-energy cosmic ray events with an
associated high multiplicity of secondary produced hadrons. Most
notably Fermi and Landau in the early 1950's initiated a statistical
approach coupled with a hydrodynamical description of the high
energy density matter, which is formed and subsequently evolves in
such prototype strong interactions \cite{Fermi,Landau} (see also
\cite{Pete} for a personal historical account and further references).
Until the present day their pioneering work has
influenced further developments in this field, even if for about
two decades there existed no hint whatsoever to the most relevant if
not fundamental degrees of freedom of high energy density matter.

The statistical approach, in particular, is motivated by considering
the large amount of initial kinetic energy which is carried into the
reaction by the colliding particles and is ultimately transformed into
the large multiplicity of secondary particles, i.e. produced matter
which finally flies apart. By a fast (on the scale of fm/c) compression
of a relativistically large amount of energy ($\gg$ 1GeV/nucleon)
into a small volume (on the scale of $\mbox{fm}^3$) one couples very
effectively to strongly interacting degrees of freedom, i.e. those
which materialize as the observed hadronic particles.
Loosely speaking, a large number of initially hidden
underlying field degrees of freedom must have been excited
during this transformation of the apparently simple (ordered) initial
state into the obviously rather complicated (disordered) final state
of the collision: It seems as if an enormous amount of ``entropy''
can be ``created'' here, which experimentally increases with energy
and complexity of the colliding hadrons or nuclei. Where does this
entropy come from which is one of the characteristic features of
multiparticle production events? How can it be interpreted
microscopically in terms of a field theory of strong interactions?

Further on, working towards a statistical theory of high energy
density matter, Hagedorn led the important intermediate step of
taking into account the information about the dynamics available in
the 1960's in the statistical bootstrap model in the form of an
exponential hadronic mass spectrum \cite{QM1,Hag}.
Here, as with all former and later
models and their phenomenologically rather successful
applications, the statistical approach immediately suggests the crucially
 simplifying assumption of local thermal (kinetic and possibly
chemical) equilibrium among the relevant constituents of matter in
very inelastic
high-energy reactions. After the advent of QCD the correct theory
of strong interactions is nowadays believed
to be firmly established with quark and gluon fields as the
fundamental degrees of freedom. Thus, corroborated by the observation
that asymptotically at short distances the running coupling
of QCD vanishes and once again guided by the equilibrium assumption,
it has been an ongoing and ever stronger effort to identify
theoretically unambiguous characteristics and find experimentally
signals of a hot and dense ``quark-gluon plasma'' of almost
non-interacting quarks and gluons. This intermediate new state of
matter should be
formed at sufficiently high energy density during hadronic
or nuclear collisions (see references \cite{QM1} and \cite{QM91},
respectively, for early and latest overviews of this field).
Until very recently, however, the equilibrium assumption
(``thermalization'') has been completely ad hoc and is still
lacking any detailed
justification or deeper understanding. Even if many global features
of high-energy reactions can be reproduced in thermal models
of varying degree of sophistication,
it has remained quite a mystery, why they work so well.

Connected to the problem of large entropy production
mentioned above, the questions arising in the context of thermalization
are: Why does the thermalization, which implies a saturation of entropy
production, proceed so fast (on the scale of fm/c)? Which are the
relevant degrees of freedom in terms
of QCD and with what kind of effective interactions (``hard''
perturbative vs. ``soft'' non-perturbative physics)
that contribute most efficiently to thermalization? How do these
``randomizing'' degrees of freedom hide again during the final
hadronization process?

Questions pertaining directly to the latter hadronization or confinement
problem presumably will have to stay off-stage for a while.
However, even those concerning (time scales of) entropy production and
thermalization cannot yet be seriously addressed. As we explain in
this paper, the necessary quantitative studies have not been clearly
defined from a conceptual point of view. It is our purpose here to
present an analysis of the entropy production in
very inelastic high-energy reactions with an eye towards the parton
picture,
to provide a plausible hypothesis about the underlying randomization
process, and to outline some calculations under simplifying
assumptions, which should strengthen the proposed scheme.

The plan of our paper is as follows. In Sec. 2 we introduce the
definitions and formal apparatus which will be applied to analyze the
entropy problem. We employ density matrices and,
especially, we explain the notions of {\it Schmidt and
pointer states in the Hilbert space of a complex system} and of
{\it environment-induced quantum decoherence}. They play an important
role in studies of the measurement process and the classical limit
of quantum theory (Ref. \cite{Zu0} provides a useful introduction
with numerous references to original work; see also the review
\cite{Om}). Recent attempts
to formulate a consistent approach to quantum cosmology initiated by
Gell-Mann and Hartle and others have given a new impetus to study these
questions again \cite{Zu0,GH}, which touch upon the foundations
of quantum theory. Basically, there one tries to understand, why the
universe considered as a closed quantum system (in an
overall pure state) looks as classical as it does.

For example, low-energy propagating modes could be coupled to
dislocalized discrete modes left over from a topological phase in
string theory, which latter ones are hidden from a low-energy observer
performing ``measurements'' by local scattering methods \cite{Ellis}.
Thus, there is a dynamical separation of modes according to whether
they are observable in the low-energy limit of string theory or not.
The unobservable modes naturally constitute an environment which may
induce quantum decoherence in the subsystem of the observable modes.
This in turn leads to {\it non-equilibrium evolution}, i.e. non-unitary
quantum mechanics in the subsystem with an associated direction of
time and entropy production, in particular.

Analogously we
may ask, why a pure-state high-energy scattering experiment lends
itself to a statistical description characterized by a large apparent
entropy and possibly quasi-classical propagation of partons. In
this more down-to-earth context of strong interactions similar
phenomena should occur on a microscopical scale ($\Lambda _{QCD}
\approx 200 \mbox{MeV}$)
which can be related to a {\it dynamical separation}
of a complex system into an {\it observable subsystem} and an
{\it unobservable environment}. Surprisingly, the subject of our present
work has not been analyzed from this appealing point of view up to now.
The paradigm of {\it confinement} has been around for about
twenty years and it seems unavoidable by now to think of unobservable
quark and gluon states or, rather, long-wavelength
field modes feeling or producing
confinement as constituting a {\it dynamically active environment
leading to quantum decoherence of the
``almost observable'' partons} (in the sense of parton-hadron duality
or deep-inelastic scattering), to which we have become used. Thus,
the situation is opposite to the one in string theory; in QCD almost
constant low-energy field modes constitute the
environment for the ``hard'' partons.

This may
have important consequences for the parton picture as applied to
nuclear collisions or complex reactions, in general, where factorization
 and assumptions about the quasi-classical propagation of partons
between successive hard scatterings come into play \cite{Klaus,Italy}.
The so-called ``sudden approximation'' underlying the parton model,
which can consistently be justified {\it within} perturbation theory,
could fail or, rather, be justified on a deeper level
precisely if and when the strongly coupled environment
becomes important. After all, the commonly referred to analogy between
a parton and a hard-struck electron in a crystal can be rather
misleading, since only in the latter case {\it all} relevant
interactions are weak.

Recently there have been
speculations that so far unexplained decohering effects on the
initial state multi-parton wave functions of colliding hadrons or
nuclei should contribute considerably to entropy production in these
reactions, see for example \cite{ML,LRT}.
We show that
the {\it mechanism for entropy production}
in strong interactions at high energy is hidden in the active role
of the environment.

In Sec. 3 we present a non-relativistic toy model of a parton
linearly coupled to gluonic oscillators which serves to illustrate
the formalism of Sec. 2 with environment-induced quantum decoherence
and entropy production in the partonic subsystem.
This model bears some resemblance to a
non-relativistic electron interacting with the quantized
electromagnetic field. The oscillator spectral density of the gluonic
environment, however, is by construction rather different and receives
its dominant contribution in the infrared, in particular.
The encouraging results obtained in a short-time strong-coupling
approximation employing the Feynman-Vernon influence functional
technique will guide us in Sec. 4, where we begin
to study the problem of strong interactions in the
parton picture.

We propose a relativistic field theory generalizing the toy
model of Sec. 3 consisting of two non-linearly but locally coupled
scalar fields, which we assume to represent the partons and the
confined (or confining vacuum) environment modes, respectively.
The purpose of this model, which may still be far removed
from a phenomenologically viable representation of QCD, is to
tackle the particular technical problems expected in the
field theory context. Loosely speaking one could also say that we
extend here the
study of {\it quantum Brownian motion} into the realm of field
theory.
As it turns out, a functional Schr\"odinger
picture approach based on the time-dependent variational principle
of Dirac seems very promising \cite{Dirac,Jackiw}. It allows
to transform the usual Heisenberg operator approach to applications
of quantum field theory, i.e. mainly time-independent scattering
problems, into a set of manageable variational equations describing the
relevant initial value problem. Furthermore, an important
advantage of this procedure here is not to be limited to the study of
weakly interacting systems, as it is the case with ordinary perturbation
 theory.

In a time-dependent Hartree-Fock type approximation we can
calculate the entropy production for {\it any} scalar parton model field
theory completely in terms of two-point Wightman or
correlation functions, which seems to
be a rather general result allowing interesting conclusions concerning
the real QCD problem. In order to learn about the time-evolution
of the two-point functions, of course, one has to specify the
interactions in detail. Similar in spirit to the BCS theory of
superconductivity, where it was realized that
particularly in the strong-coupling limit
the ``action'' is
around the Fermi surface in a conductor \cite{Anderson},
we propose here to distinguish between modes which are close to
constant fields ({\it environment modes}) and those which are not
({\it partons}).
Thus, a
Momentum Space Mode Separation is the essential feature
of our toy model field theory.

In a sequel to the present work we further evaluate the model of Sec. 4
 and explicitly
calculate the time-evolution of the parton density (functional)
matrix for various physical initial conditions.
It is important then to derive the relation with non-perturbative
(spin-, colour-, and flavour-averaged)
QCD parton structure functions \cite{Klaus,Italy,ML},
which are experimentally accessible.
Furthermore, we will refer to the
parton cascade
approach for hadronic or nuclear collisions \cite{GM}. This model,
which is based on perturbative QCD to the largest extent possible at
present and singled out among existing ``Monte Carlo'' procedures by most
 clearly specifying the model assumptions about non-perturbative soft
processes, is particularly useful as a reference frame
in trying to understand essential
aspects of entropy production in high-energy reactions.
New effects implied by a dynamical treatment of the unobservable
environment modes, as initiated here, have to be contrasted with such
a state of the art description, either to better justify or to
modify it where necessary. In particular, assumptions about the
factorization of multiple hard or semi-hard parton scatterings, i.e.
quantum mechanically independent rescattering \cite{Klaus,Italy,GM},
deserve a closer look.

Finally, we have to go over from a somewhat ad hoc model to
full QCD.
We propose an explanation of the underlying
decoherence/randomization process in terms of unobservable soft
gluon background fields, neglecting potentially important soft quark
modes (e.g. chiral condensate) for simplicity. Presumably, this
necessitates model assumptions about the properties of the background
fields, since a non-perturbative ab initio treatment of QCD seems still
out of reach. However, detailed and phenomenologically satisfactory
approximations are available here, e.g. in the form of a ``stochastic
vacuum'' model (see e.g. Refs. \cite{Dosch} and references therein).
They can be implemented
by suitably modifying the method of background field quantization
(see e.g. \cite{me} and references therein).

The studies and further remarks presented here should
indicate the next steps to verify our hypothesis about
the nature and importance of environment-induced quantum decoherence
in strong interactions, its particular relation to confinement and
unobservable long-wavelength QCD fields, and the
ensuing solution of the long-standing ``entropy puzzle'' in high-energy
reactions.

In the final Sec. 5 we discuss our main results. We also
point out further interesting applications of the formalism developed
presently. -
Throughout this paper we work with units such that
$\hbar =\mbox{c}=\mbox{k}_B=1$.

\yoursection{2. Classical Behaviour and Entropy in Pure-state
         Quantum Systems}

To begin with, we want to derive the {\it Schmidt decomposition} for
the density matrix of a closed quantum system.\footnote{The derivation
is given here to make our work selfcontained and to introduce
necessary definitions and the notation;
see also Ref. \cite{Alb} for an earlier presentation in a different
context.} - We may think of colliding hadrons or nuclei which
are sufficiently separated from any other strongly interacting matter
as an approximate realisation of such a system (neglecting any
long-range interactions) or the universe as a whole, which most
fascinating example, however, is presently understood the least. -
Thus, the complex system is described by a pure quantum state
$|\Psi\rangle$ which evolves according to some still to be specified
dynamics in an embedding Hilbert space $\cal{H}$.

The corresponding density matrix is given by a hermitian operator,
\beq
\hat{\rho}\;=\;|\Psi\rangle\langle\Psi |
\;\;, \eeq{1}
which is trivially diagonal. Suppose now that we are experimentally
limited or, rather, we are {\it forced by
the dynamics of the complex physical system to observe
only certain components of the complete state vector}
$|\Psi\rangle$. I.e., we deliberately consider a factorization of the
Hilbert space into two predetermined subspaces with orthonormal bases
$\cal{P}\equiv\{ |p\rangle\}$ (``partonic subsystem'') and
$\cal{G}\equiv\{ |g\rangle\}$ (``gluonic environment''),
respectively.\footnote{We choose these completely arbitrary labels in
purpose to prepare the reader for Secs. 3 and 4 and will be interested
particularly in the behaviour of the partonic subsystem later on.} Then,
with $\cal{H}=\cal{P}\otimes\cal{G}$,
\beq
|\Psi\rangle\; =\;\sum _{p,g}c_{pg}\; |p\rangle |g\rangle
\;\;, \eeq{2}
and,
\beq
\hat{\rho}\; =\;\sum _{p,g,p',g'}c_{pg}\;
c^{\ast}_{p'g'}\; |p\rangle |g\rangle
\langle g'|\langle p'|
\;\;, \eeq{3}
are corresponding general expansions. Furthermore, defining
\beq
|p_{\cal{G}}\rangle\;\equiv\;\sum _g c_{pg}\; |g\rangle
\;\;, \eeq{4}
we obtain
\beq
|\Psi\rangle\; =\;\sum _p |p\rangle |p_{\cal{G}}\rangle
\;\;. \eeq{5}
The linear combinations $|p_{\cal{G}}\rangle$ are not necessarily
orthogonal or normalized. However, let us introduce the density
operator for the partonic subsystem,
\beq
\hat{\rho}_{\cal{P}}\;\equiv\;\mbox{Tr}_{\cal{G}}\;\hat{\rho}\;
=\;\sum _{p,g,p'}c_{pg}\; c^\ast _{p'g}\; |p\rangle \langle p'|
\;\;, \eeq{6}
by orthonormality of the $\cal{G}$-basis. Next, we
assume that we chose the orthonormal basis of $\cal{P}$ to
be composed of the {\it eigenvectors} of $\hat{\rho}_{\cal{P}}$. Then,
by its orthonormality we find
\beq
\mbox{const}\cdot\delta _{pp'}\; =\;\langle p|\hat{\rho}_{\cal{P}}
|p'\rangle\; =\;\sum _gc_{pg}\; c^{\ast}_{p'g}\; = c_p\cdot \delta
_{pp'}
\;\;, \eeq{7}
with a positive eigenvalue $c_p\equiv \sum _g|c_{pg}|^2$.
Similarly, we obtain from eq. (\ref{4}) using eq. (\ref{7})
\beq
\langle p'_{\cal{G}}|p_{\cal{G}}\rangle \; =\;\sum _g
c_{pg}\; c^{\ast}_{p'g}\; =\;c_p\cdot \delta _{pp'}
\;\;. \eeq{8}
Thus, having chosen an eigenvector
(of $\hat{\rho}_{\cal{P}}$) basis
of $\cal{P}$, the $\{ |p_{\cal{G}}\rangle\}$ {\it must be orthogonal} as
well. Introducing the density operator of the gluonic
environment and using eq. (\ref{4}),
\beq
\hat{\rho}_{\cal{G}}\;\equiv\;\mbox{Tr}_{\cal{P}}\; \hat{\rho}\;
=\;\sum _{g,p,g'}c_{pg}\; c^{\ast}_{pg'}\; |g\rangle \langle g'|\;
=\;\sum _p|p_{\cal{G}}\rangle\langle p_{\cal{G}}|
\;\;, \eeq{9}
we also see that the $|p_{\cal{G}}\rangle$'s are the eigenvectors of
$\hat{\rho}_{\cal{G}}$ with eigenvalues $\langle p_{\cal{G}}|p_{\cal{G}}
\rangle =c_p$, by eq. (\ref{8}). Thus, surprisingly $\hat{\rho}_
{\cal{G}}$ and $\hat{\rho}_{\cal{P}}$ have identical non-zero
eigenvalues,
the number of which is determined by the dimension of the smaller of
the two subspaces $\cal{P},\cal{G}\subset\cal{H}$ or
the rank of the matrix $\{ c_{pq}\}$ in eq. (\ref{2}).
Normalizing the $|p_{\cal{G}}\rangle$'s,
\beq
|p^n_{\cal{G}}\rangle \;\equiv\;c_p^{-1/2}\; |p_{\cal{G}}\rangle
\;\;, \eeq{10}
we obtain the Schmidt decomposition \cite{Schm} of the complete state
from eq. (\ref{5}):
\beq
|\Psi\rangle\; =\;\sum _pc_p^{1/2}\; |p\rangle |p^n_{\cal{G}}\rangle
\;\;, \eeq{11}
in terms of orthonormal ``{\it Schmidt states}'' $\{ |p\rangle\}$, which
span $\cal{P}$, and $\{ |p^n_{\cal{G}}\rangle\}\subseteq\cal{G}$.
They represent the sets of {\it eigenvectors} of $\hat{\rho}_{\cal{P}}$
and $\hat{\rho}_{\cal{G}}$, respectively. Of course, the roles of
$\cal{P}$ and $\cal{G}$ can formally be exchanged in the construction.
Finally, eqs. (\ref{6},\ref{7}) yield
\beq
\hat{\rho}_{\cal{P}}\; =\;\sum _pc_p\; |p\rangle\langle p|
\;\;, \eeq{12}
and eqs. (\ref{9},\ref{10})
\beq
\hat{\rho}_{\cal{G}}\;
=\;\sum _pc_p\; |p^n_{\cal{G}}\rangle\langle p^n_{\cal{G}}|
 \;\;. \eeq{13}
Equations (\ref{11}) - (\ref{13}) are the essential results of the
{\it Schmidt decomposition} procedure.

To appreciate the importance of the Schmidt decomposition, we proceed
in several steps. First of all, we may generally assume a normalized
complex system state $|\Psi \rangle$. This implies
\beq
1\; =\;\langle\Psi |\Psi\rangle\; =\;\sum _pc_p\; =\;
\mbox{Tr} _{\cal{P}}\;\hat{\rho} _{\cal{P}}
\; =\;\mbox{Tr} _{\cal{G}}\;\hat{\rho} _{\cal{G}}
\;\;, \eeq{14}
where we used eqs. (\ref{11}) - (\ref{13}) and orthonormality
of the Schmidt states.
The sum rule (\ref{14}) suggests to interpret the positive expansion
coefficients $c_p$ in eqs. (\ref{12},\ref{13}) as {\it probabilities}
to find e.g. the partonic subsystem in the respective states
$\{ |p\rangle\}$ \cite{Feyn}. Such an interpretation is confirmed by
calculating, for example, the probability P($\tilde{p}$) to find the
partonic subsystem in a state $|\tilde{p}\rangle\in\cal{P}$,
\beq
\mbox{P}(\tilde{p})\;\equiv\;\langle\tilde{p}|
\hat{\rho}_{\cal{P}}|\tilde{p}\rangle\;=
\;\sum _pc_p\; |\langle p|\tilde{p}\rangle |^2
\;\;, \eeq{15}
which is equal to $c_{\tilde{p}}$ if $|\tilde{p}\rangle$ is one of the
Schmidt states. Or, we calculate the expectation value of an observable
$\hat{O} _{\cal{P}}$,
\beq
\langle \hat{O} _{\cal{P}}\rangle\;\equiv \;\mbox{Tr} _{\cal{P}}\;
\hat{\rho} _{\cal{P}}\;
\hat{O} _{\cal{P}}\; =\;\sum _pc_p\;\langle p|
\hat{O} _{\cal{P}}|p\rangle
\;\;, \eeq{16}
which is the expectation value of
$\hat{O} _{\cal{P}}$ in a Schmidt state times the probability
to find that state in the partonic subsystem summed over its whole
basis.

Thus, we conclude that via the Schmidt decomposition the
subsystem and its environment can be described by simultaneously
diagonal mixed state density matrices
($\hat{\rho} ^{\; 2}_{\cal{P,G}}\neq\hat{\rho} _{\cal{P,G}}$).
In particular, note the absence of quantum interference terms
$\propto\langle\tilde{p}|p\rangle\langle p'|\tilde{p}\rangle$ in
eq. (\ref{15}), for example, or $\propto\langle p'|
\hat{O} _{\cal{P}}|p\rangle$ in eq. (\ref{16}), which represent
{\it incoherent} sums. Hence, we achieved what is
called {\it quantum decoherence} in the subsystem
\cite{Zu0,Om}. Note that in distinction to eq. (\ref{16}) we obtain
\beq
\langle \hat{O} _{\cal{G}}\rangle\;\equiv \;\mbox{Tr} _{\cal{G}}\;
\hat{\rho} _{\cal{G}}\;
\hat{O} _{\cal{G}}\; =\;\sum _{p,g}c_p\;\langle g|p^n_{\cal{G}}\rangle
\langle p^n_{\cal{G}}|\hat{O} _{\cal{G}}|g\rangle
\;\;, \eeq{16.1}
where interference terms are present. In any case,
the fact remains that the complex
system {\it is} in the pure state $|\Psi\rangle$: decoherence is
achieved in the subsystem at
the expense of establishing unique {\it quantum correlations
between subsystem and environment} \cite{Zu0,Alb}, which are explicitly
 constructed in eq. (\ref{11}) above.

Next, focussing our interest on the partonic subsystem, we define its
von Neumann or {\it statistical entropy} as usual \cite{Feyn},
\beq
S_{\cal{P}}\;\equiv\; -\mbox{Tr} _{\cal{P}}\;\hat{\rho} _{\cal{P}}\;
\mbox{ln}\;\hat{\rho} _{\cal{P}}\; =\;-\sum _pc_p\;
\mbox{ln}\;c_p\; >\;0
\;\;, \eeq{17}
recalling
that $\hat{\rho} _{\cal{P}}$ is diagonal with $0<c_p<1$, see
eqs. (\ref{12},\ref{14}), and assuming always at least two non-vanishing
terms in the sum over states.
Thus, a {\bf non-zero entropy} emerges in a complex pure-state quantum
system
quite naturally, if only a subsystem is considered (or accessible to
experiment) and the trace over the remaining environment degrees of
freedom is calculated to obtain the relevant density submatrix.
Note that the total entropy vanishes,
$S\;\equiv \; -\mbox{Tr}\;\hat{\rho}\;\mbox{ln}\;\hat{\rho}\; =\;
-1\cdot\mbox{ln}\; 1\; =\; 0\;$, as it should be.
The same conclusions can be reached for the {\it linear entropy}
employed in Refs. \cite{PHZ,ZHP},
\beq
S ^{lin}_{\cal{P}}\;\equiv\;\mbox{Tr} _{\cal{P}}\;
(\hat{\rho} _{\cal{P}}-\hat{\rho} ^{\; 2}_{\cal{P}})\; =\;
1\; -\;\mbox{Tr} _{\cal{P}}\;\hat{\rho} ^{\; 2}_{\cal{P}}\; =\;
1\; -\;\sum _pc^{\; 2}_p
\;\;, \eeq{18}
which more directly measures the ``(im)purity'' of a density matrix
($0<S ^{lin}_{\cal{P}}<1$) and is often easier to handle
than the standard statistical entropy. In Sec. 4.2, eq. (\ref{110}),
we show that the linear entropy provides a {\it lower bound} on the
statistical entropy.

Finally, the evolution of the complex system has to be taken into
account. Up to now our considerations were limited to a fixed
pure state and its Schmidt decomposition which, therefore, has to be
recalculated from one instant to the next following the evolving system.
 Of course, the {\it unitary evolution} according to the Schr\"odinger
equation,
\beq
i\partial _t\; |\Psi\rangle\; =\;\hat{\mbox{H}}\; |\Psi
\rangle\;\; ,\;\;\;\; |\Psi\rangle\; =\;
\mbox{e} ^{-\textstyle{i}\hat{\mbox{H}}\textstyle{t}}\; |\Psi _0\rangle
\;\;, \eeq{19}
leaves the closed quantum system in a pure state, i.e. eq. (\ref{1})
remains valid with $|\Psi\rangle$ denoting the time-dependent state
vector.
We may split the hamiltonian,
\beq
\hat{\mbox{H}}
\;\equiv\;
\hat{\mbox{H}} _{\cal{P}}\; +\;
\hat{\mbox{H}} _{\cal{PG}}\; +\;
\hat{\mbox{H}} _{\cal{G}}
\;\;, \eeq{20}
where
$\hat{\mbox{H}} _{\cal{P\; (G)}}$ acts only on the partonic subsystem
(gluonic environment) and
$\hat{\mbox{H}} _{\cal{PG}}$ describes the interaction connecting the
two factorized subspaces introduced above, see eq. (\ref{2}).

Until the end of this section we neglect the interaction term,
$\hat{\mbox{H}} _{\cal{PG}}\approx 0$, and consider only an
{\it exactly separable system} for simplicity. Furthermore, let us
{\it assume} that a Schmidt basis of $\cal{P}$ simultaneously
diagonalizes
$\hat{\mbox{H}} _{\cal{P}}$. This implies
\beq
[\; \hat{\rho} _{\cal{P}}\; ,\;
\hat{\mbox{H}} _{\cal{P}}\; ]\; =\; 0\;\; ,\;\;\;\mbox{and}\;\;\;
\partial _t\;\hat{\rho} _{\cal{P}}\; =\; 0
\;\;. \eeq{21}
Thus, we may write
$\hat{\mbox{H}} _{\cal{P}}\; =\;\sum _pE_p\; |p\rangle\langle p|\;$,
where $E_p$ are the associated energy eigenvalues. Then, the most
general $\hat{\rho} _{\cal{P}}$ allowed by eqs. (\ref{21}) and
orthonormality of the Schmidt states is of the form given by
eq. (\ref{12}). It includes the interesting case of a ``pseudo-thermal
ensemble'',
\beq
\hat{\rho} _{\cal{P}}(\beta _p)\;\equiv\;\frac{
\mbox{exp}\{ -\sum _p\beta _pE_p\; |p\rangle\langle p|\} }{\mbox{Tr}
_{\cal{P}}\;\mbox{exp}\{ -\sum _p\beta _pE_p\; |p\rangle\langle p|\} }
\; =\;\sum _p\frac{ \mbox{e} ^{-\beta _pE_p}}{Z}\; |p\rangle\langle p|
\;\;, \eeq{22}
where $Z(\beta _p)\;\equiv\;\sum _p\mbox{exp}\{ -\beta _pE_p\}$;
for $\beta _p\rightarrow\beta$ the usual density operator of a
canonical ensemble results \cite{Feyn}. Of course, we find a non-zero
entropy here by eqs. (\ref{17}) or (\ref{18}). We remark that eqs.
(\ref{21}) by themselves imply that the expansion coefficients
(probabilities) in eqs. (\ref{12},\ref{13}) are constant,
$\partial _t\; c_p\; =\; 0$. However, from our assumption
that $\{ |p\rangle\}$ diagonalizes
$\hat{\mbox{H}} _{\cal{P}}$, we cannot further conclude
$[\; \hat{\rho} _{\cal{G}}\; ,\;
\hat{\mbox{H}} _{\cal{G}}\; ]\; =\; 0$. In general, depending
on the initial state $|\Psi _0\rangle$ in eq. (\ref{19}), the Schmidt
basis states $|p^n_{\cal{G}}\rangle\in\cal{G}$ will show a
complicated time-dependence.

Exact separability, naturally, cannot be expected to correspond to
physically interesting situations in any realistic way: (A) Consider
two decoupled field theories, one of which, the partonic subsystem,
is additionally assumed to be essentially a free theory, cf.
eqs. (\ref{21}) above; properly (anti)symmetrized
partonic multiparticle states built from single parton
momentum eigenstates constitute a suitable Schmidt basis here;
coherent states do as well, if one
neglects complications due to their overcompleteness. - Note that a
Schmidt basis is, of course, not restricted to single-particle states. -
 Or, (B) consider
a metal as a free electron gas in a background lattice, which in turn
can be described by a free phonon gas in the harmonic limit \cite{Feyn}.

At this point we anticipate that a non-vanishing {\it interaction}
$\hat{\mbox{H}} _{\cal{PG}}$ will significantly influence the
evolution of the subsystem and its environment. In Sec. 3 we present
analytical results for a rather simple interacting complex system.
It turns out
that under favourable conditions, loosely speaking in terms of the
previous example
(A), the partonic Schmidt basis remains rather stable with respect to
the evolution of the complex system and, especially, allows a parton
to be represented by a slowly evolving
gaussian wavepacket. Thus, it behaves
essentially as a {\it classical particle} under the influence of
the interaction with the gluonic environment.

This result provides part of the motivation for our attempt to
understand entropy production in strong interactions at high energy as
being related to the quantum decoherence of the relevant Schmidt
states, which would be {\it stable} in an idealized situation.
Such states are commonly referred to as ``{\it pointer
states}'' \cite{Zu0,Alb}. Typically a pointer in a {\it classical}
measuring apparatus corresponds to a preferred quantum pointer basis
being selected dynamically, such that the interactions in the complex
system (observed subsystem + environment + measuring apparatus) preserve
 the stable classical character of the pointer. Its positions on a scale
 can be described by states, the quantum superpositions of which are
dynamically suppressed (meaning decoherence). Thus, in distinction to
the well-defined Schmidt states, pointer states refer to a dynamical
situation. Up to now no general criteria exist which guarantee their
existence. Sometimes they may be realized only in an approximate way
in a given complex system \cite{Alb}, which not necessarily has to be
separable into a classically behaving subsystem plus its (quantum
correlated) environment. The purpose of Sec. 4 is to begin to study
these questions for partons in a gluonic environment.

\yoursection{3. Entropy Production via Pointer States
             in a Non-relativistic Model}

\vskip 0.2cm
\noindent
{\bf 3.1 Time-evolution of the Density Submatrix}
\vskip 0.3cm
In the following we study a complex system of a particle linearly
coupled to oscillators as a simple example
to illustrate the concepts of the
previous Section 2. For a most relevant
review of previous work on the Caldeira-Leggett model in the context
of quantum Brownian motion we refer to
Ref. \cite{Grab}, see in particular Part III. Variants of this model
have been applied before
to the quantum decoherence problem (``reduction
of the wave packet'') \cite{PHZ} and to the problem of radiation
damping for a non-relativistic electron coupled to the quantized
electromagnetic field \cite{BC}, where special attention has
been paid to the initial condition of the complex system \cite{HA}.

For our purposes the {\it Caldeira-Leggett hamiltonian} can be written
in analogy to eq. (\ref{20}) with
\beq
H_{\cal{P}}\;\equiv\;\frac{p^2}{2M}\;\; ,\;\;\;
H_{\cal{G}}\;\equiv\;\sum _{n=1}^N \left\{
\frac{p_n^2}{2m_n}\;+\;\half m_n\omega _n^2x_n^2\right\}
\;\;, \eeq{24}
describing the partonic subsystem (``parton'') by a single free
non-relativistic particle (moving in one dimension for simplicity)
and the gluonic environment by a set of $N$ harmonic oscillators, and
with the linear coupling given by
\beq
H_{\cal{PG}}\;\equiv\; -x\sum _{n=1}^{N}c_n\left\{
x_n-\frac{c_n}{2m_n\omega _n^2}x\right\}
\;\;. \eeq{25}
Thus,
\beq
H\; =\;\frac{p^2}{2M}\; +\;\sum _{n=1}^N\left\{
\frac{p_n^2}{2m_n}\; +\;\half m_n\omega _n^2\left (x_n-
\frac{\textstyle{c_n}}{\textstyle{m_n
\omega _n^2}}x\right )^2\right\}
\;\;. \eeq{26}
Choosing $c_n\equiv m_n\omega _n^2$, we see that the model becomes
explicitly translationally invariant \cite{Grab,BC,HA}, which
presently also motivates
the introduction of the parton ``potential renormalization'' term
$\propto x^2$ in eq. (\ref{25}); it can be visualized as a ``{\it parton
moving along with its gluonic springs attached}''.
In this form the model still
allows to describe a large variety of physically interesting systems
by choosing an appropriate spectral density for the environment
\cite{Grab},
\beq
I(\omega )\;\equiv\;\sum _{n=1}^N\frac{c_n^2}{2m_n\omega _n}
\;\delta (\omega -\omega _n)\;=\;\frac{1}{2}\sum _{n=1}^Nm_n\omega _n^3
\;\delta (\omega -\omega _n)
\;\;. \eeq{27}
As is well known, the environment induces noise and dissipation
in the subsystem, which are essentially determined by $I(\omega )$
(and initial conditions). Assuming a quasi-continuous
distribution of environmental oscillators, spectral densities
$I(\omega )\propto\omega ^k$ (for sufficiently small $\omega$)
with $k>0$ have
been widely studied before, e.g. \cite{PHZ,Grab,HA}. They correspond to
so-called sub-Ohmic ($k<1$), Ohmic ($k=1$), and supra-Ohmic ($k>1$)
environments, respectively. - The electron-radiation field system
involves a supra-Ohmic environment with $k=3$ \cite{BC}.

Here we consider a general spectral density in terms of an arbitrary
dimensionless function $F$,
\beq
I(\omega )\;\equiv\; g\Omega ^3F(\Omega ^{-1}\omega )
\;\Theta (\Omega -\omega )
\;\;, \eeq{28}
where $g$ is a dimensionless coupling constant and
$\Omega$ denotes a high-frequency cutoff.
For later purposes we introduce the (zero temperature)
``noise'' and ``dissipation'' kernels \cite{PHZ,Grab}, $\nu$ and $\eta$,
\beq
\nu (s)\;\equiv\;\int _0^\infty d\omega\; I(\omega )\;\mbox{cos}
(\omega s)\;\; ,\;\;\;\eta (s)\;\equiv\; -\int _0^\infty d\omega
\; I(\omega )\;\mbox{sin}\; (\omega s)
\;\;; \eeq{29}
see, for example, eqs. (\ref{40},\ref{41}) below for their dynamical
effects.
In the {\it short-time limit}, $\Omega s\ll1$, we obtain:
\beq
\nu (s)\; =\; g_\nu\Omega ^4[1\; +\;\mbox{O} (\Omega ^2s^2)]
\;\; ,\;\;\;
\eta (s)\; =\;-g_\eta\Omega ^4[\Omega s\; +\;\mbox{O} (\Omega ^3s^3)]
\;\;, \eeq{30}
with $g_\nu\equiv g\int _0^1dx\; F(x)$ and $g_\eta\equiv
g\int _0^1dx\;xF(x)$. This limit may be completely irrelevant for
macroscopic quantum systems. However, assuming $\Omega$ to be vaguely
related to a separation of non-perturbative strong interactions
with the gluonic environment from a perturbative regime at a scale on
the order of $\Lambda _{QCD}\approx 200$ MeV
and relevant time scales considerably less than 1 fm, we expect
that something analogous to
the short-time limit will be of interest there.

Having specified our toy model as above, our aim now is to
calculate the density matrix $\rho _{\cal P}$ of the partonic
subsystem, see eq.(\ref{6}), and to study the consequences of its
time evolution. To achieve this, the dependence on the gluonic
environment (harmonic oscillator degrees of freedom) has to be
integrated out. In the Caldeira-Leggett model this can be done
exactly using the Feynman-Vernon influence functional technique
\cite{Feyn,Grab}, since the coordinates (and momenta) appear at most
quadratically in the environmental hamiltonian and in the interaction,
cf. eqs. (\ref{24},\ref{25}).

Assuming the initial condition that the total density matrix
factorizes,
\beq
\rho (0)\;\equiv\;\rho _{\cal P}(0)\cdot\rho _{\cal G}(0)
\;\;, \eeq{31}
where $\rho _{\cal P}(0)$ describes the initial state of the parton,
which will be defined below, and $\rho _{\cal G}(0)$ denotes the
density matrix of the gluonic environment which is assumed to be in
the ground state at $t_0=0$; i.e., the environmental oscillators only
perform zero-point motion (``{\it vacuum fluctuations}'') initially.
Then, the parton density matrix at a later time $t$ is calculated
with the appropriate propagator,
\beq
\rho _{\cal P}(x_f,x_f',t)\; =\;\int dx_i\; dx_i'\;
J(x_f,x_f',t;x_i,x_i',0)\;
\rho _{\cal P}(x_i,x_i',0)
\;\;, \eeq{32}
which has the path integral representation \cite{Grab},
\beq
J(x_f,x_f',t;x_i,x_i',0)\; =\;\frac{1}{Z}\int {\cal D}q\; {\cal D}q'\;
\mbox{e} ^{\textstyle{i(S_{\cal P}[q]-S_{\cal P}[q'])}}\;
\mbox{e} ^{\textstyle{-\Phi [q,q']}}
\;\;, \eeq{33}
with boundary conditions $q(0)\equiv x_i$, $q(t)\equiv x_f$, and
for $q'$ analogously. Here $Z$ is a normalization factor, which will
conveniently be fixed at the end of the calculation, in order to
preserve the normalization of $\rho _{\cal P}$ or, equivalently,
of the parton wave function; $S_{\cal P}$ denotes the free parton
action corresponding to $H_{\cal P}$, eq. (\ref{24}),
\beq
S_{\cal P}[q]\;\equiv\;\int _0^tds\; \half M\dot{q} ^2
\;\;, \eeq{34}
and $\Phi$ is the Feynman-Vernon influence functional describing the
influence of the environment on the subsystem. In the present case
it is completely determined by the noise and dissipation kernels
\cite{Grab}, eqs. (\ref{29}),
\beqar
\Phi [q,q']&=&\int _0^tds\int _0^sdu\;
[q(s)-q'(s)]\left\{\nu (s-u)[q(u)-q'(u)]
+i\eta (s-u)[q(u)+q'(u)]\right\} \nonumber \\
&\;&+\;\Phi _{loc}[q,q']
\;\;, \eeqar{35}
where the ``localizing'' part of the influence functional,
\beq
\Phi _{loc}[q,q']\;\equiv\;i\sum _{n=1}^N\frac{c_n^2}{2m_n\omega _n^2}
\int _0^tds\; [q^2(s)-q'^2(s)]
\;\;, \eeq{36}
comes from the term $\propto x^2$ in eq. (\ref{25}). The constant
appearing in eq. (\ref{36}) can be calculated,
\beq
\frac{1}{2}M\omega _0^2\;\equiv\;
\sum _{n=1}^N\frac{\textstyle{c_n^2}}{\textstyle{2m_n\omega _n^2}}
\; =\;\frac{1}{2}\sum _{n=1}^Nm_n\omega _n^2\; =\;
\int _0^\infty \frac{\textstyle{d\omega}}{\textstyle{\omega}}\;
I(\omega )\; =\;g_0\Omega ^3
\;\;, \eeq{37}
where we used eqs. (\ref{27},\ref{28}) at intermediate steps and
$g_0\equiv g\int _0^1dx\; x^{-1}F(x)$.
Equation (\ref{37}) shows that $I(\omega )$ has to vanish sufficiently
fast in the infrared, in order to avoid a
divergence here.

In any case,
we observe that the resulting path integrals are gaussian and, hence,
can be evaluated exactly.
Introducing new coordinates (with a unit Jacobian),
$y\equiv q-q'$ and $z\equiv \half (q+q')$, one obtains from eq.
(\ref{33}):
\beq
J(y_f,z_f,t;y_i,z_i,0)\; =\;\frac{1}{Z}\int {\cal D}y\; {\cal D}z\;
\mbox{e} ^{\textstyle{i\Sigma [y,z]}}
\;\;, \eeq{38}
with the boundary conditions $y(0)\equiv y_i=x_i-x_i'$,
$z(0)\equiv z_i=\half (x_i+x_i')$, and $y(t)\equiv y_f=x_f-x_f'$,
$z(t)\equiv z_f=\half (x_f+x_f')$, and where the relevant total
action now is given by
\beq
\Sigma [y,z]\;\equiv\;\int _0^tds\left\{
M\dot{y}\dot{z}\; -\;M\omega _0^2yz
\; +\;\int _0^sdu\; y(s)[i\nu (s-u)y(u)
-2\eta (s-u)z(u)]\right\}
\;\;. \eeq{39}
Since $\nu$ is an even function, the second to last term here may be
rewritten such that both integrations range from 0 to $t$.

The calculation of the path integrals proceeds in the standard way
\cite{Feyn,Grab}. Since they are gaussian, the essential coordinate
dependence of the propagator can be obtained from the total action,
eq. (\ref{39}), evaluated with the extremal (``classical'')
paths.\footnote{An overall time-dependent factor stemming from the
remaining integrations over gaussian fluctuations around the classical
paths is absorbed into the normalization factor $Z$, see eq. (\ref{33}).}
 They are described by the usual Euler-Lagrange equations. Separate
variations w.r.t. $y$ and $z$, respectively, yield the equations of
motion,
\beq
M\ddot{z} (s)+M\omega _0^2z
+2\int _0^sdu\;\eta (s-u)z(u)\; =\;i\int _0^tdu\;
\nu (s-u)y(u)
\;\;, \eeq{40}
\beq
M\ddot{y} (s)+M\omega _0^2y
-2\int _s^tdu\;\eta (s-u)y(u)\; =\;0
\;\;. \eeq{41}
Using the transformation $\tilde{y} (t-s)\equiv y(s)$, one finds that
eq. (\ref{41}) written in terms of $\tilde{y} (\tilde{s}\equiv t-s)$
formally coincides with the homogeneous part of eq.
(\ref{40}). Thus, starting at $s=t$, it describes a corresponding
motion backwards in time, however, with different boundary conditions.
Since the equations are linear, real and imaginary parts of the
solution for $z$ can be calculated separately ($y$ is real by eq.
(\ref{41}) and the boundary conditions). Fortunately, it turns out
that the imaginary part of $z$ does not contribute to the minimal
action \cite{Grab,HA} and, therefore, is irrelevant at present.

Then,
we only have to solve one equation with the structure of the
homogeneous real part of eq. (\ref{40}) with appropriate boundary
conditions. Its general solution can be easily found in the
short-time limit, cf. eqs. (\ref{30}), with the Laplace transform
technique,
\beq
z(s)=\frac{\textstyle{1}}{\textstyle{f_-^2+f_+^2}}\left\{
z(0)[f_-^2\cosh (f_-s)+f_+^2\cos (f_+s)]
+\dot{z}(0)
[f_-\sinh (f_-s)+f_+\sin (f_+s)]\right\}
, \eeq{42}
where $f_\pm\equiv\Omega\{\pm g_0\Omega /M+[(g_0\Omega /M)^2+
2g_\eta \Omega /M]^{1/2}\}^{1/2}$ and $g_0$ as introduced in eq.
(\ref{37}). Employing
the transformation mentioned after eq. (\ref{41}) and inserting the
correct boundary conditions from eq. (\ref{38}), we obtain:
\beqar
z(s)&=&z_i[g_-\cosh (s_-)+g_+\cos (s_+)] \nonumber \\
&\;&+\left\{ z_f-z_i[g_-\cosh (t_-)+g_+\cos (t_+)]\right\}
\frac{f_-\sinh (s_-)+f_+\sin (s_+)}{f_-\sinh (t_-)+f_+\sin (t_+)}
\;\;, \label{43} \\
y(s)&=&y_f[g_-\cosh (t_--s_-)+g_+\cos (t_+-s_+)]
\nonumber \\
&\;&+\left\{ y_i-y_f[g_-\cosh (t_-)+g_+\cos (t_+)]\right\}
\frac{f_-\sinh (t_--s_-)+f_+\sin (t_+-s_+)}
{f_-\sinh (t_-)+f_+\sin (t_+)}\;\;, \nonumber \\
&\;&
\eeqar{44}
with $s_\pm\equiv f_\pm s$, $t_\pm\equiv f_\pm t$ and $g_\pm\equiv
f_\pm ^2(f_-^2+f_+^2)^{-1}$. Note
that in the short-time {\it weak-coupling limit} $0\leq s_\pm\leq t_\pm
\ll 1$. In this limit the results could be further simplified.

However, in the short-time limit generally
we have to use
the full expressions for the classical trajectories from eqs.
(\ref{43},\ref{44}), when calculating the minimal action. As mentioned
above, only the real part of $z$ contributes to it. Thus, after a
partial integration of the first term on the r.h.s. of eq. (\ref{39})
and making use of the real part of eq. (\ref{40}), the minimal action
is given by
\beqar
\Sigma _{min}(y_f,z_f,t;y_i,z_i,0)&=&
M\left (y_f\dot{z} _f-y_i\dot{z} _i\right )+
\frac{1}{2}i\int _0^tds\int _0^tdu\;\nu(s-u)y(s)y(u)
\nonumber \\
&=&M\left (y_f\dot{z} _f-y_i\dot{z} _i\right )+
\frac{1}{2}ig_\nu\Omega ^4\left [\int _0^tds\;y(s)\right ]^2
\;\;, \eeqar{45}
using (\ref{30}) in the last step and where $\dot{z} _{i,f}\equiv
\dot{z} (s=0,t)$, as determined by eq. (\ref{43}). Inserting eqs.
(\ref{43},\ref{44}), the final result is:
\beq
\Sigma _{min}(y_f,z_f,t;y_i,z_i,0)\; =\;\Sigma _R\; +\;i\Sigma _I
\;\;, \eeq{46}
in terms of
the real and imaginary parts of the minimal action,
\beqar
\Sigma _R&=&\frac{\textstyle{M}}
{\textstyle{f_-\sinh (t_-)+f_+\sin (t_+)}}
\left\{ [y_fz_f+y_iz_i]
[f_-^2\cosh (t_-)+f_+^2\cos (t_+)]\right . \nonumber \\
&-&y_iz_f[f_-^2+f_+^2]\nonumber \\
&-&\left . y_fz_i[g_-f_-^2+g_+f_+^2+
2g_-f_+^2\cosh (t_-)\cos (t_+)+(g_+-g_-)f_+f_-\sinh (t_-)
\sin (t_+)]\right\} \nonumber \\ [1ex]
&\equiv&ay_fz_f+by_iz_i+cy_fz_i+dy_iz_f
\;\;, \eeqar{47}
\beqar
\Sigma _I&=&\frac{\textstyle{\half g_\nu\Omega ^4}}
{\textstyle{[f_-\sinh (t_-)+f_+\sin (t_+)]^2}}
\left\{y_i[\cosh (t_-)-\cos (t_+)]\right . \nonumber \\ &\;&+y_f[
\frac{\textstyle{2g_-f_+^2}}{\textstyle{f_-f_+}}
\left . \sinh (t_-)\sin (t_+)+(g_+-g_-)(1-\cosh (t_-)\cos (t_+))]
\right\}^2 \nonumber \\ [1ex]
&\equiv&Ay_i^2+By_iy_f+Cy_f^2
\;\;, \eeqar{48}
and where we introduced the abbreviations for the time-dependent
coefficients for later convenience.
We remark that for the non-interacting case, i.e.
$g_0=g_\nu =g_\eta =0$,
the imaginary part vanishes, whereas the real part
reproduces the well-known $1/t$-term,
\beq
\Sigma _R^0\; =\;\frac{\textstyle{M}}{\textstyle{t}}
\{y_fz_f+y_iz_i-y_fz_i-y_iz_f\}
\;\;, \eeq{49}
which characterizes the free particle density matrix propagator.
Collecting
the above results, eqs. (\ref{38}) and (\ref{46}) - (\ref{48}),
the propagator $J$ immediately becomes
\beq
J(y_f,z_f,t;y_i,z_i,0)\; =\;\frac{\textstyle{1}}{\textstyle{
\tilde{Z} (t)}}\;\mbox{e} ^{\textstyle{i\Sigma _R-\Sigma _I}}
\;\;. \eeq{50}
The normalization factor $\tilde{Z}$
(cf. the last footnote) will be calculated shortly.

We consider, in particular, a normalized gaussian wave packet as the
initial state of a parton with momentum $p$,
\beq
\psi (x_i,0)\;\equiv\;\langle x_i,0|p\rangle\;\equiv\;
\alpha\;\mbox{e} ^{\textstyle{-\half x_i^2/\beta ^2}}\;
\mbox{e} ^{\textstyle{ipx_i}}
\;\;, \eeq{51}
with a corresponding density matrix,
\beq
\rho _{\cal P}(x_i,x_i',0)\; =\;\langle x_i,0|p\rangle\langle p|
x_i',0\rangle\; =\;\alpha ^2\;\mbox{e} ^{\textstyle{-\half
(x_i^2+x_i'^2)/\beta ^2}}\;\mbox{e} ^{\textstyle{ip(x_i-x_i')}}
\;\;. \eeq{52}
Then, normalization initially requires
\beq
1\; =\;\mbox{Tr}_{\cal P}\;\hat{\rho} _{\cal P}\; =\;
\int _{-\infty}^{\infty}dx\;\rho _{\cal P}(x,x,0)\; =\;
\pi ^{1/2}\alpha ^2\beta
\;\;, \eeq{53}
which fixes $\alpha =\pi ^{-1/4}\beta ^{-1/2}$. The normalization,
however, has to be preserved through the time evolution of the
system. Therefore, we proceed to calculate the time-dependent
parton density matrix. It follows from eq. (\ref{32}), using
eqs. (\ref{47},\ref{48},\ref{50},\ref{52}) and the coordinate
transformation from eq. (\ref{38}) above, after doing the two
gaussian integrals,
\beq
\rho _{\cal P}(y_f,z_f,t)\; =\;\frac{\textstyle{\pi ^{1/2}}}
{\textstyle{\tilde{Z} (t)\xi}}\;\mbox{e}
^{\textstyle{\left (\half D(y_f,z_f,t)/\xi\right )^2+iay_fz_f
-\left (C+\quarter\beta ^2c^2\right )y_f^2}}
\;\;, \eeq{54}
with
\beq
D(y_f,z_f,t)\;\equiv\; idz_f+ip-(B+\half\beta ^2bc)y_f \;\; ,\;\;\;
\xi\;\equiv\; (A+\quarter\beta ^{-2}+\quarter\beta ^2b^2)^{1/2}
\;\;. \eeq{55}
We remark that
\beq
|\psi (x,t)|^2 \; =\; \rho _{\cal P}(x,x,t)\; =\;
\rho _{\cal P}(y_f=0,z_f=x,t)
\; =\; \frac{\textstyle{\pi ^{1/2}}}
{\textstyle{\tilde{Z} (t)\xi}}\;\mbox{e}
^{\textstyle{-\quarter (dx+p)^2/\xi ^2}}
\;\;, \eeq{56}
which, when integrated over $x$, fixes the normalization of
the propagator. The result is:
\beq
\tilde{Z} (t)\; =\;2\pi |d|^{-1}
\;\;, \eeq{57}
which, of course, has to come out independently of the chosen
initial state.

Several interesting limits of the above results,
which are generally
valid in the short-time approximation, might be studied. It is
instructive to evaluate the non-interacting case first of all.
Inserting the time-dependent
coefficients from eqs. (\ref{47},\ref{48}),
we obtain from eqs. (\ref{56},{57})
in the limit $g\rightarrow 0$:
\beq
\rho _{\cal P}^0(y_f=0,z_f=x,t)\; =\;\pi ^{-1/2}w_0^{-1}(t)\;
\mbox{e} ^{\textstyle{-(x-v_0t)^2/w_0^2(t)}}
\;\;, \eeq{58}
\beq
w_0(t)\;\equiv\; (\beta ^2+\beta ^{-2}M^{-2}t^2)^{1/2}\;\; ,\;\;\;
v_0\;\equiv\;\frac{\textstyle{p}}{\textstyle{M}}
\;\;, \eeq{59}
which shows the well-known spreading of the wave packet due to the
time-dependent width $w_0$ and its motion with velocity $v_0$
according to the classical law. By comparison with eqs.
(\ref{56},\ref{57}) we find the corresponding width and velocity for
the interacting system,
\beq
w(t)\;\equiv\; 2\xi |d|^{-1}\;\; ,\;\;\; v(t)\;\equiv\; -pd^{-1}
t^{-1}
\;\;. \eeq{60}
Thus, using eqs. (\ref{54}) - (\ref{57}) and (\ref{60}), the full density
 matrix becomes
\beqar
\rho _{\cal P}(y_f,z_f,t)&=&\pi ^{-1/2}w^{-1}(t)\;
\mbox{e} ^{\textstyle{-\left (z_f-v(t)t\right )^2/w^2(t)}}
\nonumber \\
&\;&\cdot\;
\mbox{e} ^{\textstyle{-y_f^2\left (C+\quarter\beta ^2c^2-d^{-2}
[B+\half\beta ^2bc]^2/w^2(t)\right )}} \nonumber \\
&\;&\cdot\;\mbox{e} ^{\textstyle{iy_f\left (az_f-2d^{-1}
[B+\half\beta ^2bc](z_f-v(t)t)/w^2(t)\right )}}
\;\;, \eeqar{61}
which for $g\rightarrow 0$ reduces to
\beqar
\rho _{\cal P}^0(y_f,z_f,t)&=&\rho _{\cal P}^0(y_f=0,z_f,t)\;\cdot\;
\mbox{e} ^{\textstyle{-\quarter y_f^2\beta ^2M^2t^{-2}\left [
1-\beta ^2/w_0^2(t)\right ]}} \nonumber \\
&\;&\cdot\;\mbox{e} ^{\textstyle{iy_fMt^{-1}\left (z_f[1-\beta ^2/
w_0^2(t)]+v_0t\beta ^2/w_0^2(t)\right )}}
\;\;, \eeqar{62}
cf. eqs. (\ref{58},\ref{59}).

\vskip 0.35cm
\noindent
{\bf 3.2 Quantum Decoherence in the Short-time Strong-coupling Limit}
\vskip 0.3cm
Next, we turn to a discussion of the {\it short-time strong-coupling
limit}, which we define through the following conditions:
\beqar
&1.&\Omega t\;\ll 1\;\;\;\mbox{(``short time'')}\;\;;
\nonumber \\
&2.&g_0\;\gg\; 1\;\;,\;\;\; g_0/g_{\nu ,\eta}\;\gg\; 1\;\;\;
\mbox{(``strong coupling'')}\;\;;
\nonumber \\
&3.&\Omega /M\;\gg\; 1
\;\;. \eeqar{63}
Note that the second of
conditions 2. implies that the function $F$ specifying the spectral
density $I$, eq. (\ref{28}), has to be strongly infrared
dominated by the definitions of $g_{\nu ,\eta}$ and $g_0$ following eqs.
(\ref{30}) and (\ref{37}), respectively.
Condition 3. is added mainly for convenience. From conditions 1. - 3. we
obtain
\beq
f_+^2\;\approx\; 2g_0(\Omega /M)\Omega ^2\;\;,\;\;\;
f_+\;\gg\; f_-\;\;,\;\;\; g_+\;\approx\; 1\;\;,\;\;\;
g_-\;\ll\; 1
\;\;, \eeq{64}
where we used the definitions of $f_\pm$ and $g_\pm$ following eqs.
(\ref{42}) and (\ref{44}), respectively. To be definite, we will be
interested particularly in times $t$ such that $f_+t\leq\mbox{O}(1)$.
Some approximate relations follow here for the time-dependent
coefficients defined in eqs. (\ref{47},\ref{48}),
\beqar
a\; =\; b\;\approx\; Mf_+\cot (f_+t)\;\; ,\;\;\;
c\;\approx\; d\;\approx\; -Mf_+/\sin (f_+t)\;\;, \nonumber \\[2ex]
A\;\approx\; C\;\approx\;\half B\;\approx\;\half g_\nu\Omega ^4
[1-\cos (f_+t)]^2/[f_+^2\sin ^2(f_+t)]
\;\;, \eeqar{65}
which will be useful henceforth.
Note that all of eqs. (\ref{65}) approach the correct limit for
$g\rightarrow 0$ or $t\rightarrow 0$.
Evaluating eqs. (\ref{60}) in the short-time strong-coupling
limit, using (\ref{65}), one finds
\beq
v(t)\;\approx\;\frac{\textstyle{p}}{\textstyle{M}}\;
\frac{\textstyle{\sin (f_+t)}}{\textstyle{f_+t}}
\; =\;v_0\;
\frac{\textstyle{\sin (f_+t)}}{\textstyle{f_+t}}
\;\;, \eeq{66}
\beq
w^2(t)\;\approx\;\beta ^2[\cos ^2(f_+t)+
\frac{\textstyle{g_\nu}}{\textstyle{2g_0^2(\beta\Omega)^2}}
(1-\cos (f_+t))^2]\; +\;
\beta ^{-2}M^{-2}f_+^{-2}\sin ^2(f_+t)
\;\;. \eeq{67}
Generally,
the results show several remarkable features, if one compares the
expression for the density matrix obtained from
eq. (\ref{61}) in the present limit with the non-interacting
case given in eq. (\ref{62}).

To begin with, all time-dependent functions entering eqs.
(\ref{60},\ref{61}) through the coefficients evaluated in eqs.
(\ref{65}) are governed by a single {\it dynamical
time scale} $f_+^{-1}$, cf. (\ref{64}). From previous experience we
expect that their precise form is sensitive to the initial condition
\cite{PHZ,Grab,HA}, particularly the simple factorization in eq.
(\ref{31}). The fact that in the present limit trigonometric
functions dominate over the hyperbolic ones in eqs. (\ref{47},\ref{48}),
 which usually would rather determine the long-time behaviour of the
propagator etc., also points towards this conclusion. We remark that
for $t\rightarrow 0$ eq. (\ref{61}) yields $\rho _{\cal P}\rightarrow
\rho _{\cal P}^0$, eq. (\ref{62}), in agreement with the factorized
initial condition. Furthermore, there is obviously and quite naturally
a dependence on the initial state of the parton parametrized by the
width $\beta$ and momentum $p$. Note, however, had we chosen, for
example, a gaussian superposition in momenta instead of eq. (\ref{51}),
\beq
\tilde{\psi}(x_i,0)\;\equiv\;\tilde{\alpha}\; \mbox{e}
^{\textstyle{-\half x_i^2/\beta ^2}}\int _{-\infty}^{\infty}dp'\;
\mbox{e} ^{\textstyle{ip'x_i}}\; \mbox{e} ^{\textstyle{-(p'-p)^2/
P^2}}
\;\;, \eeq{68}
this would simply replace $\xi ^2\rightarrow \xi ^2+\quarter P^2$,
cf. (\ref{55}), and leave the density matrix the same otherwise,
eq. (\ref{54}), apart from a constant overall factor. We will consider
another, more interesting parton initial state below.

Turning to the diagonal elements of the density matrix, $\rho
_{\cal P}(0,z_f,t)$ from eq. (\ref{56}) and (\ref{61}) in particular,
we observe that they are completely determined by the velocity and
width, $v$ and $w$ as obtained in eqs. (\ref{66}) and (\ref{67}),
respectively:
\vskip .15cm
\noindent
{\bf (i)} The free velocity $v_0$ is modified by a slowly decreasing
function
of time (scaled by $f_+$), which shows an influence of the gluonic
environment oscillators on the parton as an effective {\it friction}
force.
\vskip .15cm
\noindent
{\bf (ii)} Compared to the spreading of the wave packet of a free
particle,
see $w_0$ in (\ref{59}), the width has become a very slowly evolving
function of time in the region of interest to us, $f_+t\leq\mbox{O}(1)$,
and can actually be smaller than the initial width $w(0)=w_0(0)=\beta$
for a suitable choice of the parameters ({\it localization}); note that
the term $\propto g_\nu$, which is due to the noise kernel, cf.
(\ref{29}), presents an increasing contribution to $w$ which can
be neglected, if the initial width is not excessively too small.
\vskip .15cm
In fig. 1 we illustrate the above results by showing the unique
{\it deceleration function} $v/v_0$, see eq. (\ref{66}), as a function
of the dimensionless
variable $t_+\equiv f_+t=\Omega t(2g_0\Omega /M)^{1/2}$.
Furthermore, we show the relative {\it change of the width} of the parton
 wave packet, cf. eq. (\ref{67}), as a function of the same variable:
\beq
\frac{\textstyle{w(t_+)-\beta}}{\textstyle{\beta}}\; =\;
\left\{\cos ^2(t_+)+\alpha _< [1-\cos (t_+)]^2+\alpha\sin ^2(t_+)
\right\}^{1/2}-1
\;\;, \eeq{68.1}
with the parameters given by
\beq
\alpha\;\equiv\;\frac{\textstyle{1}}{\textstyle{2g_0}}
\frac{\textstyle{\Omega}}{\textstyle{M}}(\beta\Omega )^{-4}\;\;,\;\;\;
\alpha _<\;\equiv\;\alpha\cdot\frac{\textstyle{g_\nu}}{\textstyle{g_0}}
\frac{\textstyle{M}}{\textstyle{\Omega}}(\beta\Omega )^2
\;\;. \eeq{68.2}
Note that the expression corresponding to eq. (\ref{68.1}) for a
{\it free} particle ($g=0$) can be written similarly,
\beq
\frac{\textstyle{w_0(t_+)-\beta}}{\textstyle{\beta}}\; =\;
\left\{ 1+\alpha\; t_+^{\;2}
\right\}^{1/2}-1
\;\;, \eeq{68.3}
which is also shown in one case in fig. 1 for comparison.

By modifying the spectral function of our toy model, eq. (\ref{28}),
one could optimize the above results in the sense that (i) the
environmental friction is minimized and simultaneously (ii) the
stability of the parton wave packet is maximized. It is, however, a
general property of linear coupling models \`a la Caldeira-Leggett
that friction and localization go together to some extent \cite{Grab}.
We remind ourselves that for
supra-Ohmic environments with spectral exponent $k>2$, as
discussed after eq. (\ref{27}), one even finds an effectively free
quantum particle with renormalized properties in the long-time limit.
Interestingly, this applies for a non-relativistic electron coupled
to the quantized electromagnetic field, where $k=3$ \cite{BC}. Thus,
as could be expected, our parton toy model specified by eqs.
(\ref{28}) - (\ref{30}) and (\ref{63}), in particular, behaves very
different from the electron-radiation field system.

At this point we want to pause for a moment and turn our attention
back to the discussion of partonic Schmidt and pointer states in Sec. 2
before. After all, the toy model in this section is meant
to illustrate those concepts and to provide an explicit example, how
they may be realized in a physical system, which is vaguely oriented
towards QCD partons. It should have become clear that the
single-parton wave packet, eq. (\ref{51}), trivially constitutes a
{\it Schmidt basis} for the single-parton (plus gluonic environment)
system under consideration, cf. eqs. (\ref{11}) - (\ref{13}).
However, far less trivial seems the fact
that this Schmidt basis also is dynamically rather stable in
the sense of points {\bf (i)} and {\bf (ii)} above. Thus, a parton here
behaves in a good approximation as a {\it classical particle}
moving in a dissipative environment. Hence, we have shown that the
Schmidt basis defined by the chosen initial wave packet state also
presents an approximate realization of a {\it pointer state} basis.
To complete this demonstration, we have to show that these pointer
states have appropriate decoherence properties as discussed
after eq. (\ref{16}).

Thus, instead of considering the off-diagonal density matrix elements
from eq. (\ref{61}) (for $y_f\neq 0$) in detail,\footnote{
Generally, one finds with the help of eqs. (\ref{65}) an {\it increased}
 gaussian (in $y_f$) fall-off as compared to the non-interacting
case, eq. (\ref{62}), with time-dependences governed by the scale
$f_+^{-1}$ discussed above, and corresponding modifications of the
phase factor.} we now consider a different parton initial state.
Our aim is to show that
\vskip .15cm
\noindent
{\bf (iii)} the gluonic environment enforces
the evolution of a {\it pure state} consisting of a {\it coherent
superposition of pointer states} into a {\it mixed state}
consisting of a {\it decoherent sum} related to those states; i.e.,
the interference terms of the initial density matrix become
dynamically suppressed as time goes on.
\vskip .15cm
\noindent
This is the {\bf origin of
entropy production} in the present model, as we shall see.

Instead of the parton initial state defined in eq. (\ref{51}), we
want to study the coherent superposition of two such wave packets
with momenta $+\tilde{p}$ and $-\tilde{p}$:
\beq
\tilde{\psi} (x_i,0)\;\equiv\; N^{-1/2}[\psi _{+\tilde{p}}(x_i,0)
+\psi _{-\tilde{p}}(x_i,0)]\;\; ,\;\;\;
N\;\equiv\; 2[1+\mbox{e} ^{\textstyle{-(\tilde{p}\beta )^2}}]
\;\;, \eeq{69}
where the index $\pm\tilde{p}$ is introduced to indicate the
respective momenta and $N$ is the necessary additional normalization
constant, see eqs. (\ref{51}) - (\ref{53}) for comparison.
Equation (\ref{69}) yields the
corresponding initial density matrix,
\beqar
\tilde{\rho} _{\cal P}(x_i,x_i',0)&=&N^{-1}\tilde{\psi} (x_i,0)
\tilde{\psi} ^\ast(x_i',0) \nonumber \\
&=&N^{-1}[
\rho _{\cal P}(x_i,x_i',0)_{+\tilde{p}}+
\rho _{\cal P}(x_i,x_i',0)_{-\tilde{p}}+
\rho _{int}(x_i,x_i',0)]
\;\;, \eeqar{70}
with the pure state density matrix elements
$\rho _{\cal P}(x_i,x_i',0)_{\pm\tilde{p}}$ as in eq. (\ref{52})
and the interference term
\beq
\rho _{int}(x_i,x_i',0)\;\equiv\;\alpha ^2
\mbox{e} ^{\textstyle{-\half (x_i^2+x_i'^2)/\beta ^2}}
[\mbox{e} ^{\textstyle{+i\tilde{p} (x_i+x_i')}}
+\mbox{e} ^{\textstyle{-i\tilde{p} (x_i+x_i')}}]
\;\;. \eeq{71}
Then, the time evolution of $\tilde{\rho} _{\cal P}$, which follows
the linear law of eq. (\ref{32}), can be evaluated term by term. The
result for $\rho _{{\cal P}\;\pm\tilde{p}}$ is given by eq.
(\ref{61}) with the appropriate momentum inserted,
whereas the {\it interference term} at a later time $t$
is calculated to be
\beq
\rho _{int}(y_f,z_f,t)\; =\;\sum _\pm\rho _{\cal P}(y_f,z_f,t)\arrowvert
_{\textstyle{v\rightarrow 0 ,\; cy_f\rightarrow c(y_f\pm 2\tilde{p}
/c)}}
\;\;. \eeq{72}
Here $\rho _{\cal P}$ is given by eq. (\ref{61}) as well,
however, with the indicated substitutions, which can easily be
understood.
Since presently $v
\rightarrow 0$, i.e. $p=0$, there is no motion of the gaussian in
the center coordinate $z_f$, compare eq. (\ref{61}). On the other
hand, the substitutions $cy_f\rightarrow c(y_f\pm 2\tilde{p}/c)$
corresponding to the respective parts of the interference term,
eq. (\ref{71}), imply that parts of the resulting gaussian
in the relative coordinate $y_f$, see eq. (\ref{61}), ``move away''
with a velocity proportional to the relative momentum $2\tilde{p}$
of the superposed initial wave packets. There is also a similar
effect on the phase factor. Because of the complexity of the r.h.s.
of eq. (\ref{72}), it is not illuminating
to examine the behaviour of individual density matrix elements
obtained from eq. (\ref{70}) at a later time $t$ with the help of
eqs. (\ref{61}) and (\ref{72}).

Therefore, we proceed to calculate directly the entropy for the various
interesting states. In particular,
we employ the {\it linear entropy} defined in eq.
(\ref{18}) for two reasons: a) $S_{\cal P}^{lin}$ has a simple bound
from above, which eventually gives a good qualitative impression of the
saturation of entropy production; b) practically, for any somewhat
complicated non-diagonal density matrix,
it seems impossible to calculate the standard entropy $S_{\cal P}$
according to eq. (\ref{17}) without performing a more or less difficult
diagonalization first of all.

Let us begin by calculating the linear entropy $S_1^{lin}$ for the
simple pointer state (\ref{51}) with its density matrix (\ref{61})
using eq. (\ref{18}):
\beqar
S_1^{lin}&=&1-\mbox{Tr} _{\cal P}\;\hat{\rho} _{\cal P}^{\;2}\; =\;
1-\int _{-\infty}^{\infty}dy_f\int _{-\infty}^{\infty}dz_f\;
\rho _{\cal P}(y_f,z_f,t)\;\rho _{\cal P}(-y_f,z_f,t) \nonumber \\
&=&1-\half c_1^{-1/2}w^{-1}
\;\;, \eeqar{73}
\beq
c_1\;\equiv\;C+\quarter\beta ^2c^2-d^{-2}w^{-2}[B+\half\beta ^2bc]^2
\;\;, \eeq{74}
independently of the parton momentum $p$.
In the non-interacting limit ($g\rightarrow 0$), where the previously
defined time-dependent functions reduce to $A=B=C=0$, $a=b=-c=-d=M/t$,
and $w\rightarrow w_0$, $v\rightarrow v_0$, eqs. (\ref{59}), the r.h.s.
of eq. (\ref{73}) correctly goes to zero for {\it all times}.
This is in agreement with eq. (\ref{18}), since here
the initial pure state remains a pure state, which has zero entropy.
However, our proposed pointer state
(\ref{51}), is not perfectly stable. Consequently, we expect some
amount of entropy to be generated due to the interaction with the
gluonic environment. Indeed, in the short-time strong-coupling limit
(\ref{63}) using eqs. (\ref{65}), one obtains from
eq. (\ref{73})
\beq
S_1^{lin}(t)\;\approx\;\textstyle{\quarter}\frac{\textstyle{g_\nu}}
{\textstyle{g_0}}
\left [\frac{\textstyle{1}}{\textstyle{
2g_0(\beta\Omega )^2}}\; [1-\cos (f_+t)]^2+
\frac{\textstyle{(\beta\Omega )^2}}{\textstyle{\Omega /M}}\;
\sin ^2(f_+t)
\right ]
\;\;, \eeq{75}
which is very small for suitable $\beta\Omega =\mbox{O}(1)$ and
according to the assumptions (\ref{63}).
In this sense the pointer state (\ref{51})
conforms with the so-called {\it predictability sieve} \cite{ZHP},
which characterizes pointer states as producing the least possible
entropy increase. Furthermore, note that a perfect pointer state would
have to diagonalize the relevant part of the total hamiltonian,
$H_{\cal P}+H_{\cal PG}$, as defined in eqs. (\ref{24},\ref{25}).
We did not seriously try to improve the
wave function (\ref{51}) to optimize its pointer state character.
However, apart from the points {\bf (i)} and {\bf (ii)} discussed
above, we will find additional support for having made a reasonable
guess by its decoherence properties.

For this purpose we calculate the linear
entropy $S_2^{lin}$ for the coherent superposition of two
pointer states, eq. (\ref{69}), with its density matrix $\tilde{\rho}$
at a finite time $t$ following from eqs. (\ref{61}) and (\ref{70}) -
(\ref{72}) as described after eq. (\ref{71}).
The result of a lengthy
but straightforward calculation, which involves a number of
double gaussian integrations as in eq. (\ref{73}), can be represented
as follows:
\beqar
&&S_2^{lin}\; =\; 1-\mbox{Tr} _{\cal P}\;\tilde{\rho} _{\cal P}^
{\;2}\; =\; 1-N^{-2}\;\mbox{Tr} _{\cal P}\left\{\hat{\rho} _{{\cal P}\;
+\tilde{p}}+\hat{\rho} _{{\cal P}\;-\tilde{p}}+\hat{\rho} _{int}
\right\} ^2\; = \label{76} \\ [2ex]
&&\; 1-N^{-2}c_1^{-1/2}w^{-1}\left\{
[2\cdot\half +\mbox{e} ^{\textstyle{-2(\tilde{p} /dw)^2-\half c_6^{\;2}/
c_1}}]+[
\mbox{e} ^{\textstyle{2c_3+\half c_2^{\;2}/c_1}}+
\mbox{e} ^{\textstyle{-2(\tilde{p}\beta )^2}}] \right . \nonumber \\
&&\;\;\; +\left .[4\cos (\half c_5\tilde{p}/d+\quarter c_2c_6/c_1)
\;\mbox{e} ^{\textstyle{-\half (\tilde{p} /dw)^2+
(c_2^{\;2}+8c_3c_1-c_6^{\;2}-w^2c_5^{\;2}c_1)/8c_1}}] \right\} \;,
\nonumber \eeqar{76.1}
where we grouped terms into square brackets which come from $(\hat{\rho}
_{{\cal P}\;+\tilde{p}}+\hat{\rho} _{{\cal P}\;-\tilde{p}})^2$,
$\hat{\rho} _{int}^{\;2}$, and the cross terms involving
$\rho _{int}$ at last, respectively;
$N$ was defined in eq. (\ref{69}), $c_1$ in eq.
(\ref{74}), and the other abbreviations are
\beqar
c_2&\equiv&-c\tilde{p}\beta ^2[1-\beta ^2b^2d^{-2}w^{-2}-2Bbc^{-1}d^{-2}
w^{-2}]\;\;,\;\;\;c_3\;\equiv\; -\tilde{p} ^2\beta ^2[1-\beta ^2b^2d^{-2}
w^{-2}]\;\;, \nonumber \\
c_5&\equiv&-2\tilde{p}\beta ^2bd^{-1}w^{-2}\;\;,\;\;\;
c_6\;\equiv\; -2\tilde{p}d^{-2}w^{-2}[B+\half\beta ^2bc]
\;\;. \eeqar{77}
We should, of course, check again that in the
non-interacting limit there is no entropy produced, since the pure
state $\tilde{\psi}$, eq. (\ref{69}), which corresponds to
$\tilde{\rho}$, remains a pure state. In fact, it is instructive
to see, how this result comes about,
\beq
S_{2\;g=0}^{lin}\; =\;1-2N^{-2}\left\{ [2\cdot\half +
\mbox{e} ^{\textstyle{-2(\tilde{p}\beta )^2}}]+
[1+\mbox{e} ^{\textstyle{-2(\tilde{p}\beta )^2}}]
+[4\;\mbox{e} ^{\textstyle{-(\tilde{p}\beta )^2}}] \right\}
\; =\; 0
\;\;, \eeq{78}
where the terms in square brackets are in one-to-one correspondence
with those in
eq. (\ref{76}). We observe that all terms are exponentially small
except for three contributions: two identical terms
in the first bracket, which can be interpreted as a decoherent sum
for the two superposed pointer states, cf. eqs. (\ref{73},\ref{76});
however, there is also a large term originating from the square of
the interference term, $\hat{\rho} _{int}^{\;2}$ in eq. (\ref{76}),
which appears in the second bracket. Obviously, if there is any
considerable entropy production to occur in the interacting case, then
particularly this contribution
of the interference term has to be significantly reduced, which
essentially
amounts to the {\it decoherence effect} that we are looking for.

Using the full result, eqs. (\ref{76},{77}), evaluated in the
short-time strong-coupling limit and expanding
consistently to lowest non-trivial order in quadratically small
quantities, i.e. terms which are proportional to $g_\nu /g_0^2$ or
$g_\nu /(g_0\Omega /M)$, we find:
\beq
S_2^{lin}(t)\;\approx\; S_1^{lin}(t)-2N^{-2}\left \{
\sigma _1\;\mbox{e} ^{\textstyle{-2(\tilde{p}\beta )^2}}-\sigma _2
+4\sigma _3\;\mbox{e} ^{\textstyle{-(\tilde{p}\beta )^2}}
\right \}
\;\;, \eeq{79}
where we also used eqs. (\ref{73}) - (\ref{75}) and assumed
$\beta\Omega =\mbox{O}(1)$ again; furthermore, the terms $\propto\sigma
_{1,2,3}$ correspond to the square brackets in eqs. (\ref{76}) or
(\ref{78}) in an obvious way, with
\beqar
\sigma _1&\equiv&(\tilde{p}\beta )^2\frac{\textstyle{2g_\nu (\beta
\Omega )^2}}{\textstyle{g_0\Omega /M}}\;\sin ^2(f_+t)\;\; ,
\nonumber \\
\sigma _2&\equiv&(\tilde{p}\beta )^2\frac{\textstyle{g_\nu}}
{\textstyle{g_0^2(\beta\Omega )^2}}\; [1-\cos (f_+t)]^2\;\; ,
\nonumber \\
\sigma _3&\equiv&(\sigma _1-\sigma _2)/4
\;\;. \eeqar{80}
Several remarks are in order here. Firstly, note that all $\sigma$'s
are $\propto \tilde{p} ^2$. Thus, $\tilde{p}\rightarrow 0$ correctly
implies $S_2^{lin}(t)\rightarrow S_1^{lin}(t)$, since in this limit
we recover the simple (imperfect) pointer state (\ref{51}) with $p=0$
and its
associated entropy, see eqs. (\ref{73}) - (\ref{75}). Secondly, for
moderately large relative momentum between the two coherently
superposed pointer states in eq. (\ref{69}) as compared to their
spatial width, i.e. $\tilde{p} >\beta ^{-1}$, we obtain from eqs.
(\ref{79},\ref{80}) the final result:
\beq
S_2^{lin}(t)\;\approx\;\textstyle{\half}\sigma _2
\;\;. \eeq{81}
Thus, we conclude that an appreciable amount of linear {\it entropy}
can be produced here through the mechanism of {\it environment-induced
decoherence of coherently superposed (approximate) pointer states}.
In particular,
note that the dominant contribution $\propto \sigma _2$ comes from
the decay of the interference term, which we anticipated in
the discussion following eq. (\ref{78}). The rate is determined
by the dynamical time scale $f_+^{-1}$, cf. (\ref{64}), which also
governs the friction and localization effects mentioned in points
{\bf (i)} and {\bf (ii)} above.

It is worth noting that in the limit of very large relative momentum,
$\tilde{p}\beta\rightarrow\infty$, we find from eq. (\ref{76}) in the
short-time strong-coupling limit instead of (\ref{81}):
\beq
S_2^{lin}(t)\;\approx\; 1-\half [1-S_1^{lin}(t)][2\cdot\half]\; =\;
\half [1+S_1^{lin}(t)]
\;\;. \eeq{82}
This result corresponds simply to the expected decoherent superposition
of two equally probable states, cf. eq. (\ref{18}), which would yield
$S_{\cal P}^{lin}\; =\; 1-\{ (\half )^2+(\half )^2\}$.
Again there is a small correction due to the imperfection of our
pointer states as discussed above.

Finally, we remark that for the demonstration of entropy production or
environment-induced decohence we did not necessarily choose the
{\it optimal} superposition of approximate pointer states in the sense
of the largest or fastest effect. Our above result, eqs.
(\ref{79}) - (\ref{81}), shows that the entropy production is
larger for superposed states which are further separated in momentum,
i.e. states the wave packets of which have consequently moved further
apart in space after a given amount of time has passed. From this
observation and the fact that the interaction between partonic
subsystem and gluonic environment in our toy model, see eqs.
(\ref{25},\ref{26}), acts on the respective coordinates rather than
the momenta one may be tempted to speculate that the superposition
of initially spatially (rather than in momentum) separated pointer
states would decohere even faster and, thus, produce entropy at a
higher rate. Indeed, in the different context of Ref. \cite{PHZ},
such an effect of a {\it preferred observable} (coordinate vs.
momentum) has been found.

These remarks complete the demonstration of dynamically enforced
decoherence of coherent superpositions of pointer
states, point {\bf (iii)} above, and its relation to entropy
production for our toy model.

More general superpositions of states can, of course, be studied
similarly. In particular, a general parton initial state could be
decomposed into coherent sums over pairs of localized gaussian wave
packets (see also the discussion in Ref. \cite{PHZ}), for which the
present considerations apply again.

The extension to a
{\it multi-parton system} is straightforward only as long as we can
neglect perturbative interactions among themselves\footnote{This is
a question of the relative sizes of a non-perturbative time
scale corresponding to $f_+^{-1}$ here and a perturbative
``rescattering'' time, which has to be
taken up in the QCD context} and interactions mediated
through the self-coupling of the gluonic environment. Otherwise,
provided that a Feynman-Vernon type influence functional can still be
calculated (cf. Sec. 3.1), one has to face at least a generally quite
difficult mechanical $n$-body problem, i.e. a large set of coupled
equations replacing (\ref{40},\ref{41}). This seems to prohibit a
stepwise development of our toy model into a more realistic
phenomenological model. Therefore, we propose a fully relativistic
quantum field theory model instead in the following section, which
is motivated by the non-relativistic model studied here.
It might mimic some features of QCD and the parton model.

Generally, we expect to encounter the set of characteristic
formal problems posed by the requirements of Lorentz and local gauge
covariance as well as by the occurence of the usual divergences
of quantum field theory.
A more subtle difference, which is
due to going over from the non-relativistic Schr\"odinger equation to
proper wave equations describing the basic fields, will mostly
be hidden in our functional Schr\"odinger picture approach.

\yoursection{4. Decoherence and Entropy Production in a Scalar
             ``Parton'' Field Theory}

\vskip 0.2cm
\noindent
{\bf 4.1 Model of Momentum Space Mode Separation}
\vskip 0.3cm
Before attempting to generalize our previous considerations to QCD
gauge theory, we want to take an intermediate step here by studying a
much simpler
model of two coupled real scalar fields, $\phi _1$ and $\phi _2$,
representing partons and ``unobservable'' gluonic environment modes,
respectively. It is defined by the classical action
\beq
S\;\equiv\;\int d^4x\;\{ L_1+L_{12} +L_2 \}
\;\;, \eeq{83}
with
($g_{\mu \nu}
\equiv\mbox{diag}[1,-1,-1,-1]$)
\beq
L_j\;\equiv\;\half (\partial\phi _j)^2-v_j(\phi _j)\;\;, \;\;\; j\; =\;
1,2\;\;;\;\;\;
L_{12}\;\equiv\;-V(
\phi _1,\phi _2)
\;\;, \eeq{84}
i.e., we consider an {\it attractive interaction} of the two
scalar fields, in particular. Clearly, the meaning of the model
depends on how we define the separation of partons and environment
modes.

To begin with, let us assume that there was only {\it one} scalar
quantum field $\varphi$ representing the {\it total} system. Then,
similar to the background field method, we may split it,
$\varphi \equiv \phi _1+\phi _2$, according to the following
considerations.

The essential feature of QCD to be incorporated here phenomenologically
is the {\it running coupling constant}. ``Soft'' interactions
with small four-momentum transfer (squared, $|Q|^2$) have a strong
effective coupling and, generally, cannot be described in perturbation
theory. Conversely, ``hard'' processes with large $|Q|^2$ are
reliably accounted for by perturbation theory and constitute the
{\it only} basis for the comparison of QCD theory with experiment
\cite{AMuller}. Field modes which are sufficiently close to constant
field configurations and {\it
would} interact only among themselves form a sector of the total
Hilbert space which {\it would} be completely unobservable. They can
neither hadronize nor can any large-$|Q|^2$ process occur {\it within}
this sector due to the restriction of the four-momenta. Roughly,
we assume $k_0^{\;2}<m_\pi^{\;2}$ and $\vec{k}^2<\Lambda _{QCD}^2$.
Of course, there is no sharp boundary of this sort in QCD. In any case,
this defines the {\it gluonic environment} described by
the field $\phi _2$ and $L_2$ in eqs. (\ref{83},\ref{84}) and the
region II in fig. 2. (We don't distinguish soft quark
modes presently.) The complement of this part of the total Hilbert
space, region I in fig. 2, is described by $\phi _1$ and $L_1$ and
constitutes the {\it parton subsystem}, which is almost directly
observable by hadronization or via large-$|Q|^2$ processes. Finally,
we describe the interactions between partons and environment by
$L_{12}$ in eqs. (\ref{83},\ref{84}).

Several qualifying remarks are in order here: \\
1. The Momentum Space Mode Separation (``MSMS'') as introduced above
is {\it not} Lorentz invariant. Therefore, it only makes sense in a
certain frame. It should be the one in which we formulate the initial
conditions for the time-evolution problem of our field theory (cf.
Sec. 4.4). \\
2. In a non-abelian gauge theory MSMS
is {\it not} gauge invariant. Therefore, it only makes sense,
if the gauge is completely fixed; particularly, if the theory is
described in terms of physical degrees of freedom. \\
3. At present $\phi _1$ and $\phi _2$ are {\it quantum} fields,
whereas usually in the background field method the background
consists of a classical field, which does not propagate in quantum
loops, whereas the quantum field only lives on internal lines in
Feynman diagrams by construction. Obviously, a separation of field
modes according to some dynamical criterion (e.g. MSMS)
requires a non-trivial modification of the
standard background (gauge) field method \cite{me}. \\
4. There is an analogy to the non-relativistic model of
Sec. 3; the interaction $L_{12}$
may be depicted as the ``{\it partons dragging along the
unobservable background field modes}'' or vice versa. A close analogy
would be a non-local interaction involving a
form factor,
$$\int d^4x\; \tilde{L}_{12}\;\equiv\;\int d^4x\; d^4x'\;
[\phi _1(x)-\phi _2 (x)]\;\omega ^2(x-x')\;
[\phi _1(x')-\phi _2 (x')]\;\;,$$
which is, however, local in momentum space and, thus, would rather
directly generalize the interaction in eq. (\ref{26}). Unfortunately,
such an effective interaction presents difficulties,
when attempting the usual
equal-time quantization of the model. Therefore, it is also
not obvious how to write down a functional Schr\"odinger equation for
it, cf. eq. (\ref{85}) below, and we will not consider it any further.

In order to specify our model, we consider the potentials in
eqs. (\ref{84}) in more detail according to MSMS. Renormalizability
and stability require the potentials to be fourth order polynomials.
Then, a minimal version of the model may be defined by:
\beqar
v_1(\phi _1)\;\equiv\;\frac{\textstyle{1}}{\textstyle{4!}}\lambda _1
\phi _1^{\;4}&,&\;
v_2(\phi _2)\;\equiv\;-\frac{\textstyle{1}}{\textstyle{2}}\mu ^2\phi _2
^{\;2}+\frac{\textstyle{1}}{\textstyle{4!}}\lambda _2
\phi _2^{\;4} \;\;, \nonumber \\ [2ex]
V(\phi _1,\phi _2)&\equiv&\lambda _{12}\phi _1^{\;2}\phi _2^{\;2}
\;\;, \eeqar{84.1}
with the restriction $0<\lambda _1<\lambda _{12}<\lambda _2$, which is
supposed to mimic in a very schematic way the $|Q|^2$-dependence of
the strong coupling. In $v_2(\phi _2)$ we included a negative ``mass''
term to allow for a condensation of environment modes
(generating different masses for $\phi _1$ and $\phi _2$). The
$\phi _2$-potential is fully consistent with MSMS.

However, in the mutual
coupling we encounter the problem that it admits, for example, the
annihilation of two space-like
partons into two environment gluons with arbitrarily
small $|Q|^2$. This contradicts a clear-cut MSMS, such that
the respective size of couplings in (\ref{84.1}) corresponds
to the $|Q|^2$-dependent situation in QCD. The same effect
would arise in other couplings, which we presently did not include for
simplicity, and
cannot be avoided in such a crude model. Similarly, the
$\phi _1^{\;4}$-interaction admits some low-$|Q|^2$ processes, which
it should not. Therefore, only on the average (in multiple scattering)
may we hope to mimic a bit of the QCD case. This will be studied in
the sequel to the present work. There we may want to include also
$\phi _1\phi _2^{\;2}$- and
$\phi _1^{\;2}\phi _2$-couplings. They correspond to decay/fusion
and soft scattering processes, which are important in parton
cascades \cite{Klaus,Italy,GM}. In the time-dependent Hartree-Fock
approximation considered below the tree level Feynman diagrams
for the two-point functions
corresponding to the latter cubic couplings would simply carry one
classical
field $\langle\phi _{1,2}\rangle$ on one leg, whereas $V(\phi _1,
\phi _2)$ as in (\ref{84.1}) describes effective mass insertions
$\langle\phi _i^{\;2}\rangle _{i=1,2}$ and something like a
formfactor $\langle\phi _1\phi _2\rangle$, which we discussed above
and which we consider most interesting to start with.

Generally, the self-interactions $v_j(\phi _j)$ and the mutual
coupling $V(\phi _1,\phi _2)$
necessitate a {\it renormalization} of our model. It is well known, how
to carry out the standard program for scalar fields. Furthermore, it has
been extended to the study of initial-value problems recently
\cite{CM}, which would be relevant in the following.
However, since it does not
illuminate our main topic here, we simply assume for the moment a
regularization of
the arising divergences (cf. below) by a high-momentum cut-off.
In this way we also avoid a potential triviality problem (in four
dimensions) of the quartic interactions.

To proceed in analogy with Sec. 3, we consider the {\it time-evolution}
problem for the quantum field theory defined by eqs. (\ref{83},\ref{84})
 starting with given initial field configurations. In general, this
can only be done in some approximation, e.g. ordinary perturbation
theory for sufficiently small couplings in connection with any
formalism able to handle time-dependent (non-equilibrium) systems
such as (semi-classical) transport theory \cite{us} or the
Schwinger-Keldysh formalism (see e.g. Ref. \cite{CH} and references
therein). Since the interesting results of Sec. 3 depend crucially on
the short-time {\it strong-coupling} limit, cf. Sec. 3.2 and
(\ref{63}), in particular, and we anticipate the strong coupling at
small four-momentum transfer in QCD to be most important for the
environment-induced decoherence effects, we need a calculational
scheme which allows us to consider the corresponding limit here.

\vskip 0.35cm
\noindent
{\bf 4.2 Variational Approach to Time-evolution in Quantum Field Theory}
\vskip 0.3cm
It seems most efficient to employ the time-dependent variational
principle of Dirac \cite{Dirac,Jackiw} which is equivalent to the
{\it functional Schr\"odinger equation} describing the full dynamics
of a generic field $\varphi$ in the Schr\"odinger picture,
\beq
i\partial _t\Psi [\varphi;t]\; =\; H[\hat{\pi},\varphi ]\Psi [\varphi ;
t]\;\equiv\;\int d^dx\left \{-\half\frac{\textstyle{\delta ^2}}
{\textstyle{\delta\varphi ^2}}+\half (\nabla\varphi )^2+{\cal V}(\varphi
)\right \}\Psi [\varphi ;t]
\;\;, \eeq{85}
where $\Psi [\varphi ;t]\equiv\langle\varphi |\Psi (t)\rangle$ denotes
the wave functional in the $\varphi$-representation, which
corresponds to a wave function $\psi (x,t)
\equiv\langle x|\psi (t)\rangle$ for a one-dimensional
quantum-mechanical system, and $\hat{\pi} =-i\delta /\delta\varphi$
represents the canonical momentum operator conjugate to the field
(``coordinate'') $\varphi$. The dynamics is determined by the
hamiltonian $H$. In this context the completeness
and inner product
relation, respectively, involve functional integrals instead of
ordinary ones (orthogonality needs a $\delta$-functional),
\beq
\langle\Psi _1(t)|\Psi _2(t)\rangle\;\equiv\;
\int {\cal D}\varphi\;\langle\Psi _1(t)|\varphi\rangle\langle\varphi |
\Psi _2(t)\rangle\; =\;\int {\cal D}\varphi\;\Psi _1^\ast [\varphi ;t]
\Psi _2[\varphi ;t]
\;\;, \eeq{86}
which induces
\beq
\langle\Psi (t)|O(\hat{\pi} ,\varphi )|\Psi (t)\rangle\; =\;
\int {\cal D}\varphi\;\Psi ^\ast [\varphi ;t]\;
O(-i\frac{\textstyle{\delta
}}{\textstyle{\delta\varphi}},\varphi )\;\Psi [\varphi ;t]
\;\;, \eeq{87}
i.e. the evaluation of expectation values of functions of operators
corresponding to the usual quantum-mechanical formula
$$\langle\psi (t)|O(\hat{\pi} ,x]|\psi (t)\rangle\; =\;\int dx\;
\psi ^\ast (x,t)\; O(-i\frac{d}{dx} ,x)\;\psi (x,t)\;\;.$$
Furthermore, note that eq. (\ref{87}) can be rewritten in terms of a
density functional operator $\hat{\rho}$,
\beq
\langle\Psi (t)|O(\hat{\pi} ,\hat{\varphi})|\Psi (t)\rangle\; =\;
\int {\cal D}\varphi\;\langle\varphi |\Psi (t)\rangle\langle\Psi (t)|
O(\hat{\pi} ,\varphi )|\varphi\rangle\;\equiv\;\mbox{Tr}\;\hat{\rho}(t)
\; O(\hat{\pi} ,\hat{\varphi})
\;\;, \eeq{88}
which is again analogous to the usual result, cf. eqs.
(\ref{1},\ref{16},\ref{16.1}), for example. Finally, we may state the
{\it variational principle} \cite{CM,CPS},
\beqar
\frac{\textstyle{\delta\Gamma [\Psi ]}}{\textstyle{\delta\Psi}}&=&0
\;\;,\;\;\;\mbox{for all $\Psi$ with}\;\;\;\langle\Psi (t)|\Psi (t)
\rangle\; =\; 1\;\;, \label{89} \\ [2ex]
\mbox{and}\;\;\;\Gamma [\Psi ]&\equiv&\int dt\;\langle\Psi (t)|
[i\partial _t-H]|\Psi (t)\rangle
\;\;, \eeqar{90}
i.e. requiring the {\it effective action} $\Gamma$ defined in eq.
(\ref{90}) to be stationary against arbitrary variations of the
normalized wave functional $\Psi$, which vanish at $t\rightarrow\pm
\infty$, is equivalent to the exact functional
Schr\"odinger equation, eq. (\ref{85}) above. With the variational
principle in hand, eqs. (\ref{89},\ref{90}), one can solve the
time-evolution problem in quantum field theory approximately by
restricting
the variation of the wave functional $\Psi$ to a subspace of the full
Hilbert space, which means by choosing a suitably parametrized
trial wave functional.

In the following we choose to work with most general {\it gaussian
trial wave functionals}. For the generic field $\varphi$ it is defined
by
\beq
\Psi _G[\varphi ;t]\;\equiv\;N(t)\;\exp \left\{
-[\varphi -\bar{\varphi}(t)]
\left [\quarter G^{-1}(t)
-i\Sigma (t)\right ]
[\varphi -\bar{\varphi}(t)]\; +\; i
\bar{\pi}(t)
[\varphi -\bar{\varphi}(t)] \right\}
\;, \eeq{91}
where here and henceforth we use a shorthand notation suppressing
integrations over spatial variables, whenever they are obvious.
For example,
\beqar
\varphi \;\bf G^{-1}(t)\;
\bar{\varphi}(t)&\equiv&
\int d^dx\; d^dy\;
\varphi (x)\; G^{-1}(x,y,t)\;
\bar{\varphi}(y,t) \;\;, \nonumber \\ [2ex]
\bar{\pi}(t)\;
\varphi &\equiv&\int d^dx\;
\bar{\pi}(x,t)\;
\varphi (x) \;\;, \;\;\;\mbox{and}\;\;\;\mbox{tr}\;\Sigma (t)\;
\equiv\;\int d^dx\;\Sigma (x,x,t)
\;\;. \eeqar{92}
The normalization factor $N$ can be easily calculated (for symmetric
and positive-definite $G$) according to eq. (\ref{86}),
\beq
1\; =\;\int {\cal D}\varphi\;\Psi _G^\ast [\varphi ;t]
\Psi _G[\varphi ;t]\;\;\longrightarrow\;\;\;
N(t)\; =\; ({\cal N}\;\mbox{det}\; G(t))^{-1/4}
\;\;, \eeq{93}
which is a useful result for further calculations employing $\Psi _G$
(${\cal N}$ is an infinite constant which can be omitted in
the following).
The meaning of the variational parameter functions $\bar{\varphi}$,
$\bar{\pi}$, $G$, and $\Sigma$ follows from
\cite{CM,CPS}:
\beqar
&\;&\langle \varphi (x)\rangle _G\; =\;\bar{\varphi} (x,t)\;\;, \;\;\;
\langle -i\frac{\textstyle{\delta}}{\textstyle{\delta\varphi (x)}}
\rangle _G\; =\;\bar{\pi} (x,t)\;\;, \label{94} \\ [2ex]
&\;&\langle \varphi (x)\varphi (y)\rangle _G\; =\;\bar{\varphi} (x,t)
\bar{\varphi} (y,t)+G(x,y,t)\;\;, \label{95} \\ [2ex]
&\;&\langle i\partial _t\rangle _G\; =\;\bar{\pi}(t)\dot{\bar{\varphi}}
(t)-\mbox{tr}\; [\dot{\Sigma} (t)G(t)]
\;\;, \eeqar{96}
where the operator expectation values $\langle\ldots\rangle _G$ are
calculated according to eq. (\ref{87}) with $\Psi _G$. (There is a
trivial error common to Refs. \cite{CM,CPS} in the sign of the term
$\propto\bar{\pi}$ in the definition of the trial wave functional.)
Thus, the trial wave functional $\Psi _G$, eq. (\ref{91}), is a gaussian
centered at $\bar{\varphi}$ with a width $G$; $\bar{\pi}$ and
$\Sigma$ are ``conjugate momenta'' for $\bar{\varphi}$ and
$G$ \footnote{This is obvious after a partial integration of the last
term on the r.h.s. of eq. (\ref{96}), which can be
performed in the evaluation of the effective action, see eq. (\ref{97}),
 for example.}, respectively. We anticipate that the application of
a gaussian trial wave functional amounts to a description of the
field theory under consideration in terms of {\it coupled equations for
one- and two-point Wightman
functions} (``TDHF'', time-dependent Hartree-Fock
approximation). The equivalence with the Cornwall-Jackiw-Tomboulis
generating functional (effective action) for two-particle irreducible
graphs was demonstrated by those authors in Ref. \cite{Corn} for
energy eigenstates of the field.

Next, we evaluate the effective action, eq. (\ref{90}), with the
trial wave functional, eq. (\ref{91}), to obtain:
\beqar
\Gamma [\Psi _G]&=&\int dt\left\{\;
[\bar{\pi}\dot{\bar{\varphi}}-\half\bar{\pi}^2
 -\half (\nabla\bar{\varphi})^2-{\cal V}(\bar{\varphi})]
\right . \nonumber \\ [2ex]
&\;&+\;\mbox{tr}\left [\Sigma\dot{G}-2\Sigma G\Sigma-
\frac{1}{8}G^{-1}+\frac{1}{2}\nabla ^2G\right ]-\frac{\textstyle{1}}
{\textstyle{2!}}\int d^3x\; {\cal V}^{(2)}(\bar{\varphi})G(x,x)
\nonumber \\ [2ex]
&\;&-\frac{\textstyle{3}}{\textstyle{4!}}{\cal V}^{(4)}(\bar{\varphi})
\int \left . d^3x\; G^2(x,x)\;\right\}
\;\;, \eeqar{97}
where we suppressed spatial integrations as in (\ref{92}) where possible
and the time-dependence everywhere, $\nabla ^2$ acts only on either
one of the two arguments of $G$, ${\cal V}^{(n)}(\bar{\varphi})\equiv
d^n{\cal V}(\bar{\varphi})/d\bar{\varphi} ^n$, and we assume the
potential to be at most quartic. Note
that the calculation of $\langle {\cal V}(\varphi )\rangle _G$, in
particular, proceeds conveniently through a Taylor expansion at an
intermediate step:
\beq
\langle \Psi _G|{\cal V}(\varphi )|\Psi _G\rangle\; =\; N^2\int
{\cal D}\varphi\;\left\{
{\cal V}(\bar{\varphi})+\frac{\textstyle{1}}{\textstyle{2!}}
{\cal V}^{(2)}(\bar{\varphi})\varphi ^2
+\frac{\textstyle{1}}{\textstyle{4!}}
{\cal V}^{(4)}(\bar{\varphi})\varphi ^4\right\}
\mbox{e}^{-\half\varphi G^{-1}\varphi}
\;\;, \eeq{98}
where we used eqs. (\ref{87},\ref{91}) and observe that only even
powers in $\varphi$ contribute. Employing eq. (\ref{93}), one obtains
the simple gaussian integrals needed here by a suitable functional
differentiation w.r.t. $G^{-1}$ (cf. also eqs. (\ref{116},\ref{117})
below),
\beq
\left (\varphi ^{2[4]}(x)\right )_G\;\equiv\; N^2\int {\cal D}\varphi \;
\varphi ^{2[4]}(x)\;\mbox{e} ^{-\half\varphi G^{-1}\varphi}
\; =\; 1[3]\cdot G^{1[2]}(x,x)
\;\;, \eeq{99}
and, thus,
produces the potential contributions to the above result, eq. (\ref{97}).
 We remark that the terms in the first line on the r.h.s. of eq.
(\ref{97}) present the usual classical action, cf. the
hamiltonian in eq. (\ref{85}), whereas the terms in the second and
last line are O($\hbar$) and O($\hbar ^2$) quantum corrections,
respectively. From the effective action the relevant coupled equations
of motion are obtained by independent variations w.r.t. to the
variational parameter functions $\bar{\varphi}$, $\bar{\pi}$, $G$,
and $\Sigma$. This will be performed for our model in the following.

\vskip 0.35cm
\noindent
{\bf 4.3 Model-independent Entropy in TDHF Approximation}
\vskip 0.3cm
After the above short review and formal preparations we are
finally ready to study the {\it scalar parton field theory}
defined in
eqs. (\ref{83},\ref{84}) with the variational technique.
We assume a most general gaussian trial wave functional in
product form,
\beqar
\Psi _{12}[\phi _1,\phi _2;t]&\equiv&N_{12}(t)\;\Psi _{G_1}[\phi _1;t]\;
\Psi _{G_2}[\phi _2;t]
\nonumber \\ [2ex]
&\;&\cdot\;
\exp\left\{-\half [\phi _1-\bar{\phi} _1(t)]\left [G_{12}(t)-i
\Sigma _{12}(t)\right ]
[\phi _2-\bar{\phi} _2(t)]\right\}
\;, \eeqar{100}
with the normalized gaussians on the r.h.s. as defined in eq. (\ref{91}).
 Here, however, each of the variational parameter functions in
the expression for $\Psi _{G_j}[\phi _j;t]$
carries an index $j$ and $N_{12}$ denotes an additional normalization
factor. The latter is necessary, since we included here an essential
exponential describing possible {\it two-point correlations} between the
parton field $\phi _1$ and the gluonic environment field $\phi _2$. It
is obtained from the normalization condition, cf. eq. (\ref{93}),
\beqar
1&=&\int {\cal D}\phi _1{\cal D}\phi _2\;\Psi _{12}^\ast [\phi _1,\phi _2
;t]
\Psi _{12}[\phi _1,\phi _2;t] \nonumber \\ [2ex]
&=&\left (N_1N_2N_{12}\right )^2
\int {\cal D}\phi _1{\cal D}\phi _2\;
\exp\left\{-\half [\phi _1G_1^{-1}\phi _1+\phi _2\{G_2^{-1}-
G_{12}G_1G_{12}\}\phi _2]\right\} \nonumber \\ [2ex]
&=&\left (N_1N_2N_{12}\right )^2
\;\left (\det\; G_1^{-1}\right )^{-1/2}\left (\det\{G_2^{-1}-
G_{12}G_1G_{12}\}\right )^{-1/2} \nonumber \\ [2ex]
&=&
N_{12}^{\;2}
\left (\det\{ 1-G_{12}G_1G_{12}G_2\}\right )^{-1/2} \nonumber \\ [2ex]
&\longrightarrow&\;\; N_{12}(t)\; =\;
\left (\det\{ 1-G_1(t)G_{12}(t)G_2(t)G_{12}(t)\}\right )^{1/4}
\;\;, \eeqar{101}
where we suitably shifted the fields to reach the second equation,
assume $G_{1,2}$ to be {\it symmetric and positive definite} and
similarly have to restrict the argument of the determinant in the
final result to assure the existence of the functional integrals
(dropping irrelevant constants as mentioned after eq.
(\ref{93})). We remark that the ``1'' in eq. (\ref{101}) represents
a $\delta$-function (of the space coordinates), which will occur
frequently in the following. Furthermore, note that $N_{12}$ is
symmetric under $G_1\leftrightarrow G_2$, as it should be.

To appreciate the importance of the non-diagonal factor
(in $\phi _{1,2}$) in eq. (\ref{100}), which involves parton-environment
 correlations, we calculate the {\it partonic density functional}
according to the general definition discussed in Sec. 2, cf. eq.
(\ref{6}):
\beq
\hat{\rho} _{\cal P}(t)\;\equiv\;\mbox{Tr} _{\cal G}\;\hat{\rho} (t)\;
 =\;\mbox{Tr} _2\; |\Psi _{12}(t)\rangle\langle\Psi _{12}(t)|
\;\;, \eeq{102}
which yields the ``matrix elements''
\beqar
\langle \phi _1|\hat{\rho} _{\cal P}(t)|\phi _1'\rangle &\equiv &
\rho _{\cal P}[\phi _1,\phi _1';t]\; =\;
\int {\cal D}\phi _2\;\Psi _{12}^\ast [\phi _1',\phi _2;t]
\Psi _{12}[\phi _1,\phi _2;t] \label{103} \\ [2ex]
&=&N_{12}^{\;2}(t)\;\Psi _{G_1}^\ast [\phi _1';t]\Psi _{G_1}[\phi _1
;t] \nonumber \\ [1ex] &\;&\cdot
\int {\cal D}\phi _2\;\Psi _{G_2}^\ast [\phi _2;t]
\Psi _{G_2}[\phi _2;t]\;\exp\left\{ -X_1[\phi _1',\phi _1;t](\phi _2
-\bar{\phi} _2)\right\}  \nonumber \\ [2ex]
&=&N_{12}^{\;2}(t)\;\Psi _{G_1}^\ast [\phi _1';t]\Psi _{G_1}[\phi _1
;t]\;\exp\left\{\half
X_1[\phi _1',\phi _1;t]G_2(t)
X_1[\phi _1',\phi _1;t]\right\}
\;\;, \nonumber \eeqar{103.1}
with
\beq
X_1[\phi _1',\phi _1;t]\;\equiv\;
\half [\phi _1'-\bar{\phi}_1(t)]\left [G_{12}(t)+i\Sigma _{12}(t)\right ]
+
\half [\phi _1 -\bar{\phi}_1(t)]\left [G_{12}(t)-i\Sigma _{12}(t)\right ]
\;, \eeq{104}
and where we used eq. (\ref{100}), see also eqs. (\ref{88},\ref{91}).
In addition to $G_{12}$, cf. eq. (\ref{101}), we now also assume
$\Sigma _{12}$ to be {\it symmetric}, such that $X_1^{\;t}=X_1$ w.r.t.
space coordinates.
Obviously, {\it if the correlation functions} $G_{12}(t)$
{\it and} $\Sigma _{12}(t)$ {\it vanish},
then the partonic density functional $\rho _{\cal P}$ is just
the one of an independent scalar field in the {\it pure state}
$|\Psi _{G_1}(t)\rangle$. According to our previous general
considerations
in Sec. 2, cf. also the application in Sec. 3.2, the partonic
entropy has to vanish in this case.

We can verify this conclusion by a direct
calculation (similar to the one above) in the case
of the {\it linear entropy}, which was defined in eq. (\ref{18}), using
the parton density functional from eqs. (\ref{103},\ref{104}):
\beq
S_{\cal P}^{lin}\; =\; 1\; -\;\mbox{Tr} _{\cal P}\;
\hat{\rho}_{\cal P}^{\;2}\; =\;
1\; -\; N_{12}^{\;4}\;\det (A)^{-1/2}\det (A-BA^{-1}B)^{-1/2}
\;\;, \eeq{105}
where again, as in $N_{12}$ from eq. (\ref{101}),
we have to assume that the arguments of the determinants
are symmetric and {\it positive definite}. They are
defined in terms of
\beqar
A(t)&\equiv &1-\half G_1(t)G_{12}(t)G_2(t)G_{12}(t)+\half G_1(t)\Sigma _
{12}(t)G_2(t)\Sigma _{12}(t) \;\;, \nonumber \\ [1ex]
B(t)&\equiv&\half G_1(t)G_{12}(t)G_2(t)G_{12}(t)+\half
G_1(t)\Sigma _{12}(t)G_2(t)\Sigma _{12}(t)
\;\;. \eeqar{106}
The above result can be further simplified, if we assume all two-point
functions to be {\it translation invariant}. Then, after
a Fourier transformation, the quantities appearing in the arguments of
the determinants in eq. (\ref{105}) can be handled like ordinary
numbers, which yields:
\beqar
S_{\cal P}^{lin}(t)&=&1\; -\;\det\;\left (\frac{\textstyle{1-
G_1G_{12}G_2G_{12}}}{\textstyle{1+G_1\Sigma _{12}G_2\Sigma _{12}}}
\right )^{1/2}\;=\;1-\exp\{\half\;\mbox{Tr}\;\ln (\ldots )\}
\nonumber \\ [2ex]
&=&1-\exp\{\frac{1}{2}\mbox{V}_d\int
\frac{\textstyle{d^dk}}{\textstyle{(2\pi )^d}}
\;\ln\left (\frac{\textstyle{1-G_1(k,t)G_2(k,t)G_{12}^{\;2}(k,t)}}
{\textstyle{1+G_1(k,t)G_2(k,t)\Sigma _{12}^{\;2}(k,t)}}\right )\}
\;\;, \eeqar{107}
where $\mbox{V}_d\equiv\int d^dx$ in $d$ dimensions.

Note that for any
translation invariant
two-point function $F$ here, with
\beq
F(x,y;t)\; =\; F(x-y;t)\;\equiv\;\int
\frac{\textstyle{d^dk}}{\textstyle{(2\pi )^d}}\;\mbox{e} ^{ik\cdot
(x-y)}\; F(k;t)
\;\;, \eeq{108}
the following constraints,
\beq
F(k;t)\; =\; F(-k;t)\;\;, \;\;\;\mbox{and} \;\;
F(k;t)\; >\; 0
\;\;, \eeq{109}
are implied by
the requirements of being symmetric and positive definite.

Before we continue, we want to draw attention to several remarkable
features of the above result for the linear entropy, which seem to be
of a rather general nature:
\vskip .15cm
\noindent
{\bf I.} Neither the ``mean fields'' $\bar{\phi} _{1,2}$, nor their
conjugate momenta $\bar{\pi} _{1,2}$, nor the imaginary parts
$\Sigma _{1,2}$ of the parton and gluonic environment two-point
functions, respectively,
contribute to the entropy, i.e.
to Tr$\;\hat{\rho} _{\cal P}^{\;2}$. This result is formally true for
{\it arbitrary powers},
Tr$\;\hat{\rho} _{\cal P}^{\;n}$, since the corresponding terms
always cancel between successive factors of $\hat{\rho}_{\cal P}$,
when evaluating the trace. Thus, we anticipate it to hold for the
statistical entropy as well (cf. below).
\vskip .15cm
\noindent
{\bf II.} As expected, the {\it entropy vanishes}
for {\it vanishing correlations
between partons and gluonic environment}, i.e. $G_{12}(t)=\Sigma _
{12}(t)=0$ (independent subsystems).
\vskip .15cm
\noindent
{\bf III.} If either one of the {\it widths} of the parton or environment
 wave functionals vanishes,
$G_1(t)\rightarrow 0$ or $G_2(t)\rightarrow 0$
(cf. eqs. (\ref{91},\ref{100}),
then the entropy vanishes. Thus, if either subsystem is
constrained to follow essentially the {\it classical equations of
motion}, then there is
{\it no entropy production} (cf. discussion at the end of Sec. 4.4).
\vskip .15cm
\noindent
{\bf IV.} The above conclusions {\bf I.} - {\bf III.} and the result
for the linear entropy
as given in eq. (\ref{107}) are {\it model-independent}.
They hold for {\it any} scalar field theory of partons
coupled to gluonic environment modes independently of the specific
form of the (renormalizable) interactions. They are, however, obtained
here in the {\it TDHF
approximation} as embodied in the wave functional of eq. (\ref{100}).
\vskip .15cm
\noindent
If we want to learn in detail about the time-evolution of the
system, i.e. of the variational parameter functions,
then we have to specify the interactions. The dynamics of the system
with the model interactions from Sec. 4.1 will be studied in
Sec. 4.4.

It is worthwhile stressing the importance of parton-environment
correlations one more time, since these are precisely the
quantum correlations discussed after eq. (\ref{16.1}) in
Sec. 2. Thus, our considerations presently confirm the
general idea that {\it quantum decoherence} in a subsystem,
and therewith a measurable {\it entropy production} therein, is
induced by interactions with a dynamically active
environment.\footnote{Note that a {\it functional diagonalization}
of the parton density functional, see eqs. (\ref{102}) - (\ref{104}),
would amount to an explicit construction of the time-dependent
{\it pointer states} (see Sec. 2) for the parton field.}
The parton model serves as a study case par excellence, since {\it
confinement}
dynamically separates the total system into an almost directly
observable hard parton subsystem plus an unobservable non-perturbative
environment of soft gluon modes (cf. Sec. 4.1).

Finally, we want to present here also an approximate result for the
{\it statistical
entropy}, which will later be more closely related to observables,
the particle multiplicity in high-energy reactions, in particular.
With the general definition of the statistical entropy in eq. (\ref{17})
we find,
\beqar
S_{\cal P}&=&-\sum _p\;c_p\;\ln\;c_p\;\equiv\langle -\ln\;c_p\rangle
\;\geq\; -\ln\;\langle c_p\rangle \nonumber \\ [2ex]
&=&-\ln\;\sum _p\; c_p^{\;2}\; =\; -\ln\;\mbox{Tr}_{\cal P}\;
\hat{\rho}_{\cal P}^{\;2}\; =\; -\ln\; (1-S_{\cal P}^{lin})
\;\;, \eeqar{110}
where we also used eq. (\ref{18}) and the inequality results from
Jensen's inequality for convex functions, which is the basis for
many variational principles (see e.g. Ref. \cite{Feyn}). The equality
sign holds, if the (sub-)system is in a pure state (vanishing entropy)
or if all states are equally probable, $c_p=const$, which is the case
for a thermalized system at asymptotically high temperature
($T\rightarrow\infty$), cf. eq. (\ref{22}) for example. Applying
eq. (\ref{110}) to eq. (\ref{107}), we obtain:
\beq
S_{\cal P}(t)\;\geq\;
-\frac{1}{2}\mbox{V}_d\int
\frac{\textstyle{d^dk}}{\textstyle{(2\pi )^d}}
\;\ln\left (\frac{\textstyle{1-G_1(k,t)G_2(k,t)G_{12}^{\;2}(k,t)}}
{\textstyle{1+G_1(k,t)G_2(k,t)\Sigma _{12}^{\;2}(k,t)}}\right )
\;\;. \eeq{111}
The right hand side here should provide a useful approximation for the
statistical entropy in our scalar model, if we are interested
particularly in the
initial phase of entropy production, i.e. the decoherence process
in a strong interaction at high energy, starting from a pure
quantum state.

\vskip 0.35cm
\noindent
{\bf 4.4 Effective Action and Equations of Motion}
\vskip 0.3cm

Our aim in this concluding section is to derive the effective action
for the complex system of partons and gluonic environment from which
we then obtain the relevant TDHF equations of motion, which describe the
time-evolution of the system, the partonic subsystem in particular.

First of all, it helps somewhat to simplify the following calculations,
 if we rewrite the partonic density matrix from eqs.
(\ref{103},\ref{104}). We find:
\beqar
\langle \phi _1|\hat{\rho} _{\cal P}(t)|\phi _1'\rangle &\equiv &
\rho _{\cal P}[\phi _1,\phi _1';t]
\label{112} \\ [2ex]
&=&\tilde{\Psi} _{G_1}^\ast [\phi _1';t]\tilde{\Psi} _{G_1}[\phi _1
;t]\;\exp\left\{
Y_1^\ast [\phi _1';t]G_2(t)
Y_1[\phi _1,;t]\right\}
\;\;, \nonumber \eeqar{112.1}
with
\beq
Y_1[\phi ;t]\;\equiv\;
 \half [\phi -\bar{\phi}_1]\left [G_{12}(t)-i\Sigma _{12}(t)\right ]
\;\;, \eeq{113}
and where the {\it effective gaussian} $\tilde{\Psi}_{G_1}$ is
defined as before, however, with the following replacements:
\beqar
N_1(t)&\longrightarrow&\tilde{N}_1(t)\;\equiv\;
N_1(t)N_{12}(t)\;\;, \nonumber \\ [1ex]
G_1^{-1}(t)&\longrightarrow&\tilde{G}_1^{-1}(t)\;\equiv\;
G_1^{-1}(t)A(t)\;\;, \nonumber \\ [1ex]
\Sigma _1(t)&\longrightarrow&\tilde{\Sigma}_1(t)\;\equiv\;
\Sigma _1(t)-\eighth [
\Sigma _{12}(t)G_2(t)G_{12}(t)+G_{12}(t)G_2(t)\Sigma _{12}(t)]
\;, \eeqar{114}
see eq. (\ref{91}) for the definition of the generic $\Psi _G$ and
(\ref{106}) for $A(t)$. (Recall that products of two-point
functions involve an integration over intermediate coordinates.)
Not surprisingly, the changes presently induced in the parton
density functional by the environment bear some distant similarity
with what happened to the single-parton density matrix in Sec. 3.1,
cf. eqs. (\ref{51}) - (\ref{61}). It is also worthwhile to appreciate
the analogies and differences between the Feynman-Vernon approach for
the non-relativistic toy model, eqs. (\ref{31}) - (\ref{35}), and the
TDHF
result for {\it any} two coupled scalar fields, eqs. (\ref{112}) -
(\ref{114}) above.

Next, we want to calculate the analogues of eqs. (\ref{94}) -
(\ref{96}) to see the changes caused by the environment interacting
with the scalar parton field:
\beqar
\langle \phi _1(x)\rangle&\equiv&
\int {\cal D}\phi _1\;\phi _1(x)\;\rho _{\cal P}[\phi _1,\phi _1;t]
\; =\;\bar{\phi} _1(x,t) \;\;, \label{115} \\ [2ex]
\langle -i\frac{\textstyle{\delta}}{\textstyle{\delta\phi _1(x)}}
\rangle&\equiv&\int {\cal D}\phi _1'\;{\cal D}\phi _1\;
\delta [\phi _1'-\phi _1]\; (
-i)\frac{\textstyle{\delta}}{\textstyle{\delta\phi _1(x)}}\;
\rho _{\cal P}[\phi _1,\phi _1';t]\; =\;\bar{\pi}_1(x,t)
\;\;. \nonumber \eeqar{115.1}
Thus, there is no change in the mean fields here, as expected. However,
\beqar
\langle\phi _1(x)\phi _1(y)\rangle&\equiv&
\int {\cal D}\phi _1\;\phi _1(x)\phi _1(y)
\;\rho _{\cal P}[\phi _1,\phi _1;t] \nonumber \\ [2ex]
&=&(N_1N_{12})^2\int {\cal D}\phi _1\;[\phi _1(x)+\bar{\phi} _1(x,t)]
[\phi _1(y)+\bar{\phi} _1(y,t)] \nonumber \\ [1ex]
&\;&\;\;\;\;\;\;\;\;\;\;\;\;\;\;
\;\;\;\;\;\;\;\;\;\;\;\;\;\;\cdot\;\exp\left\{-\half\phi _1G_1^{-1}[A-B]
\phi _1\right\} \nonumber \\ [2ex]
&=&
\bar{\phi} _1(x,t)\bar{\phi} _1(y,t)]\;-\; 2(N_1N_{12})^2\;
\frac{\textstyle{\delta}}{\textstyle{\delta G_1^{-1}[A-B]|_{(x,y)}}}\;
(N_1N_{12})^{-2} \nonumber \\ [2ex]
&=&
\bar{\phi} _1(x,t)\bar{\phi} _1(y,t)]\;+\;[A(t)-B(t)]^{-1}G_1(t)|_{
(x,y)}
\;\;, \eeqar{116}
where we used eqs. (\ref{93},\ref{101}), and eqs. (\ref{106}); we
suitably shifted the field to obtain the second equation and employed
(for $M$ symmetric)
\beq
\frac{\textstyle{\delta}}{\textstyle{\delta M(x,y)}}\;\det\; M\; =\;
\frac{\textstyle{\delta}}{\textstyle{\delta M(x,y)}}\;\exp\{
\mbox{tr}\;\ln\; M\}\; =\;M^{-1}(x,y)\;\det\; M
\;\;, \eeq{117}
to obtain the final result.
Obviously, the correlations with the
environment modify the parton two-point function in a non-trivial way.
Note that the relevant factor $[A-B]^{-1}$ could be expanded in a
{\it geometric series}.
We remark that equations which are formally identical to
(\ref{115}) - (\ref{116}) hold also for the field $\phi _2$.
Due to the formal symmetry of the wave functional, eq. (\ref{100}),
they are obtained by simply exchanging the indices everywhere
(``$1\leftrightarrow 2$''), which denote the parton and gluonic
background fields, respectively. Furthermore, we obtain:
\beqar
\langle i\partial _t\rangle&\equiv&
\int {\cal D}\phi _1\; {\cal D}\phi _2\;\Psi _{12}^\ast [\phi _1,
\phi _2;t]\;(i\partial _t)\;\Psi _{12}[\phi _1,\phi _2;t]
\nonumber \\ [2ex]
&=&\left\{\bar{\pi}_1(t)\dot{\bar{\phi}}_1(t)\;-\;\mbox{tr}\left [
\dot{\Sigma}_1(t)[A(t)-B(t)]^{-1}G_1(t)\right ]\right\}\;+\;
\left\{\;``1\leftrightarrow 2\mbox{''}\;\right\}
\nonumber \\ [1ex]
&\;&-\half\mbox{tr}\left [\;\dot{\Sigma}_{12}(t)\langle [\phi _1-
\bar{\phi}_1(t)][\phi _2-\bar{\phi}_2(t)]\rangle\right ]
\;\;, \eeqar{118}
where we employed
eq. (\ref{116}).
Equation (\ref{118})
generalizes eq. (\ref{96}) by adding up separate contributions
from both fields and a new term entirely due to the interaction
between them. Consequently, a similar structure will arise in the
evaluation of the total effective action below. We also find:
\beqar
&\;&\langle [\phi_1(x)-\bar{\phi}_1(x,t)]
[\phi_2(x)-\bar{\phi}_2(x,t)]\rangle
\; =\; 2N_{12}^{-1}\frac{\textstyle{\delta}}
{\textstyle{\delta G_{12}(x,y)}}
N_{12} \nonumber \\ [2ex]
&=&
-\half\{ G_2(t)G_{12}(t)[A(t)-B(t)]^{-1}G_1(t)
+[A(t)-B(t)]^{-1}G_1(t)G_{12}(t)G_2(t)\} _{(x,y)}^S
\nonumber \\ [2ex]
&=&
-[A(t)-B(t)]^{-1}G_1(t)G_2(t)G_{12}(t)|_{(x,y)}
\;\;, \eeqar{119}
where the index $S$ indicates full symmetrization w.r.t. $G_1
\leftrightarrow G_2$ and separately $x\leftrightarrow y$. This symmetry
stems from the symmetry of $G_{12}$ and
eq. (\ref{101}), which is frequently used in the course of the
calculations here with $N_{12}^{\;4}=$det$(A-B)$. For
{\it translation invariant two-point functions} the last equality
in (\ref{119}) follows. This simplifying assumption will {\it always}
be made henceforth.

As a final ingredient for the evaluation of the effective action we
need
\beqar
&\;&\langle [\phi_1(x)-\bar{\phi}_1(x,t)]^2
[\phi_2(x)-\bar{\phi}_2(x,t)]^2\rangle \nonumber \\ [2ex]
&=&(N_1N_2N_{12})^2\left\{ 2\cdot\frac{\textstyle{\delta ^2}}
{\textstyle{\delta G_{12}(x,x)^2}}+4\frac{\textstyle{\delta ^2}}
{\textstyle{\delta G_1^{-1}(x,x)
\delta G_2^{-1}(x,x)}}\right\}(N_1N_2N_{12})^{-2}
\nonumber \\ [2ex]
&=&2\langle [\phi_1-\bar{\phi}_1][\phi_2-\bar{\phi}_2]_{(x)}
\rangle ^2+\langle [\phi_1 -\bar{\phi}_1]_{(x)}^2\rangle
\langle [\phi_2 -\bar{\phi}_2]_{(x)}^2\rangle
\nonumber \\ [1ex]
&\;&+\;2G_{12}(x,x)\cdot [A-B]^{-2}G_1^{\;2}G_2^{\;2}G_{12}|_{(x,x)}
+2G_1G_2G_{12}|_{(x,x)}\cdot [A-B]^{-2}G_1G_2G_{12}|_{(x,x)}
\nonumber \\ [1ex]
&\;&+\left\{ G_1(x,x)\cdot [A-B]^{-1}G_2|_{(x,x)}+\half G_1(x,x)\cdot
[A-B]^{-2}G_1G_2^{\;2}G_{12}^{\;2}|_{(x,x)}\right .
\nonumber \\ [1ex]
&\;&\left .\;\;\;\; +\half G_1G_2^{\;2}G_{12}^{\;2}|_{(x,x)}\cdot
[A-B]^{-2}G_1|_{(x,x)}\right\}\;+\;
\left\{\;``1\leftrightarrow 2\mbox{''}\;\right\}
\;\;, \eeqar{120}
i.e. an expectation value with all four fields at the same point
(the factor 2 in the first equation is a symmetry
factor like the factor 3 in eq. (\ref{99}));
we made use of eqs. (\ref{116},\ref{119}) to rewrite terms as
``disconnected parts'' in the first line in the second equation
and remark that the
above result would look even worse without a final
simplification due to translation invariance
(intermediate spatial integrations
are suppressed as before and the obvious time dependence is
omitted).

To calculate the {\it effective action}, eq. (\ref{90}), for the scalar
field theory defined by eqs. (\ref{83},\ref{84}) in TDHF approximation
we proceed in analogy with Sec. 4.2 and evaluate various contributions
separately:
\beq
\Gamma [\Psi _{12}]\;\equiv\;\int dt\;\langle \Psi_ {12}(t)|[i\partial
_t-H]|\Psi _{12}(t)\rangle\; =\;\int dt\left\{
\langle i\partial_t\rangle -\langle H_1\rangle -\langle H_2\rangle
-\langle V_{12}\rangle\right\}
\;\;. \eeq{121}
First of all, $\langle i\partial_t\rangle$ was evaluated already in
eq. (\ref{118}) together with eq. (\ref{119}).
Secondly, similarly as the corresponding terms
(without time-derivatives) in
eq. (\ref{97}) before, we obtain after a lengthy
calculation ($j=1,2$):
\beqar
&\;&\langle H_j\rangle\;\equiv\;
\int d^dx\;\langle\Psi_{12}(t)|[-\half\frac{\textstyle{\delta ^2}}
{\textstyle{\delta\phi_j^2}}+\half (\nabla\phi_j)^2+v_j(\phi_j)]|
\Psi_{12}(t)\rangle \nonumber \\ [2ex]
&=&
\half\bar{\pi}_j^2
+\half (\nabla\bar{\phi}_j)^2+v_j(\bar{\phi}_j)
\label{122} \\ [1ex]
&\;&-\;\mbox{tr}\left [-2\tilde{\Sigma}_j[A-B]^{-1}G_j\tilde{\Sigma}_j
-\eighth G_j^{-1}[A+B]
+\half [A-B]^{-1}\nabla ^2G_j\right ] \nonumber \\ [1ex]
&\;&+\int d^3x\; [\;\frac{\textstyle{1}}
{\textstyle{2!}}v_j^{(2)}(\bar{\phi}_j)\cdot [A-B]^{-1}G_j
|_{(x,x)}
+\frac{\textstyle{3}}{\textstyle{4!}}v_j^{(4)}(\bar{\phi}_j)\cdot
([A-B]^{-1}G_j|_{(x,x)})^2\; ]
\;\;, \nonumber \eeqar{122.1}
where the same comments as after eq. (\ref{97}) apply and we made use
of equations analogous to (\ref{98},\ref{99}) which hold for the
present case.
Furthermore, we employed eq. (\ref{116}) and recalled the definition of
$\tilde{\Sigma}_1$ in eq. (\ref{114}), while $\tilde{\Sigma}_2$
follows with the familiar exchange ``$1\leftrightarrow 2$''.
We remark that eq. (\ref{122}) correctly yields the corresponding
terms in eq. (\ref{97}) for {\it vanishing correlations} ($A=1,\;B=0$)
between the fields $\phi_1$ and $\phi_2$, i.e. for independent
subsystems.

Next, we have to calculate the contribution $\langle V_{12}
\rangle$ from the mutual interaction of the fields. Employing the
appropriate Taylor expansion for a general potential, which is
at most quartic in the fields (cf. also eqs. (\ref{98},\ref{99})),
and keeping only non-vanishing terms yields:
\beqar
&\;&\langle V_{12}\rangle\;\equiv\;
\int d^dx\;\langle\Psi_{12}(t)|V(\phi_1,\phi_2)|\Psi_{12}\rangle
\nonumber \\ [2ex]
&\;&=\;V(\bar{\phi}_1,\bar{\phi}_2)
+\frac{\textstyle{2}}{\textstyle{2!}}
V^{(1,1)}
\langle (\phi_1-\bar{\phi}_1)(\phi_2-\bar{\phi}_2)\rangle
+\frac{\textstyle{6}}{\textstyle{4!}}
V^{(2,2)}
\langle (\phi_1-\bar{\phi}_1)^2(\phi_2-\bar{\phi}_2)^2\rangle
\nonumber \\ [1ex]
&\;&\;\;\;\; +\;\{\;
\frac{\textstyle{1}}{\textstyle{2!}}
V^{(2,0)}
\langle (\phi_1-\bar{\phi}_1)^2)\rangle
+\frac{\textstyle{1}}{\textstyle{4!}}
V^{(4,0)}
\langle (\phi_1-\bar{\phi}_1)^4\rangle\;\}
\; +\;\{\; ``1\leftrightarrow 2\mbox{''}\;\}
\;\;, \eeqar{123}
where $V^{(m,n)}\equiv
\frac{d^m}{d\bar{\phi} _1^{\;m}}
\frac{d^n}{d\bar{\phi} _2^{\;n}}V(\bar{\phi} _1,\bar{\phi} _2)$
and we omitted the overall
$\int d^dx$ in the last equation.
The expectation values arising here have all been calculated before.
In particular, we recall eq. (\ref{99}), which applies for the
terms in the last line with $\varphi\rightarrow\phi_{1,2}$ and
$G^{-1}\rightarrow G_{1,2}^{-1}\cdot [A-B]$ after suitably shifting the
fields; these terms obviously generate additional contributions of a
similar (mean field dependent)
structure as the potential terms $\propto v_j^{(2),(4)}$ in
eq. (\ref{122}). The most interesting terms besides the
classical contribution in the second to last line
in eq. (\ref{123}) are obtained explicitly by inserting eqs.
(\ref{119},\ref{120}).

This completes our derivation of the
Cornwall-Jackiw-Tomboulis type {\it effective action} \cite{Corn}
for any system of {\it two coupled scalars fields} with at
most quartic interactions.
We implemented Dirac's time-dependent variational principle, eqs.
(\ref{89},\ref{90}), with the help of a most general gaussian trial
wave functional, see eq. (\ref{100}). Now,
combining eqs. (\ref{118}) - (\ref{123}), we obtain the final result
(with $\hbar$'s inserted):
\beqar
&\;&\Gamma [\Psi_{12}]\; =\;\left .\int dt\;\right\{
\sum_{j=1,2}\left\{\;
\bar{\pi}_j\dot{\bar{\phi}}_j
-\half\bar{\pi}_j^2
-\half (\nabla\bar{\phi}_j)^2-v_j(\bar{\phi}_j) \right .
\nonumber \\ [2ex]
&\;&+\;\hbar\;\mbox{tr}\; [\;
\Sigma_j\dot{\bar{G}}_j
-2\tilde{\Sigma}_j\bar{G}_j\tilde{\Sigma}_j
-\frac{1}{8}G_j^{-1}[A+B]
+\frac{1}{2}\nabla ^2\bar{G}_j\; ]
\nonumber \\ [2ex]
&\;&-\;\frac{\textstyle{\hbar}}
{\textstyle{2!}}\langle\{ v_j^{(2)}
+V_j^{(2)}\}\rangle\;\mbox{tr}\; \bar{G}_j
-\frac{\hbar^2}{\mbox{V}_d}
\frac{\textstyle{3}}{\textstyle{4!}}\{ v_j^{(4)}
\left . +V_j^{(4)}\}(\;\mbox{tr}\;\bar{G}_j)^2
\;\right\}
\nonumber \\ [2ex]
&\;&+\;
\frac{\hbar}{2}\;\mbox{tr}\; [\;\dot{\Sigma}_{12}\bar{G}_1\bar{G}_2
\bar{G}_{12}\; ]
-V(\bar{\phi}_1,\bar{\phi}_2)
+\hbar
\frac{\textstyle{2}}{\textstyle{2!}}
\langle V^{(1,1)}\rangle\;
\mbox{tr}\; [\bar{G}_1\bar{G}_2
\bar{G}_{12}]
\nonumber \\ [2ex]
&\;&-\;\frac{\hbar^2}{\mbox{V}_d}
\frac{\textstyle{6}}{\textstyle{4!}}
V^{(2,2)}\left [\; 2(\;\mbox{tr}\;\bar{G}_1\bar{G}_2\bar{G}_{12})^2
+\;\mbox{tr}\;\bar{G}_1\;\mbox{tr}\;\bar{G}_2 \right .
\nonumber \\ [2ex]
&\;&\;\;\;\; +2\;\mbox{tr}\; G_{12}
\;\mbox{tr}\; [G_1^{\;2}G_2^{\;2}G_{12}[A-B]^{-2}]
+2\;\mbox{tr}\; [G_1G_2G_{12}]\;\mbox{tr}\; [G_1
G_2G_{12}[A-B]^{-2}]
\nonumber \\ [2ex]
&\;&\;\;\;\;
+\{\;\mbox{tr}\; G_1\;\mbox{tr}\;\bar{G}_2+\frac{1}{2}\;\mbox{tr}\;G_1
\;\mbox{tr}\;[G_1G_2^{\;2}G_{12}^{\;2}[A-B]^{-2}]
\nonumber \\ [2ex]
&\;&\;\;\;\;\;\;\; \left .
+\frac{1}{2}\left.\mbox{tr}\; [G_1G_2^{\;2}
G_{12}^{\;2}]\;\mbox{tr}\; [G_1[A-B]^{-2}]\;\}\; +\;
\{\; ``1\leftrightarrow 2\mbox{''}\;\}
\;\right ] \right\}
\;\;, \eeqar{124}
with $v_j^{(n)}\equiv d^nv_j(\bar{\phi}_j)/d\bar{\phi}_j^{\;n}$,
$V_j^{(n)}\equiv
d^nV(\bar{\phi}_1,\bar{\phi}_2)/d\bar{\phi}_j^{\;n}$, and
$V^{(m,n)}$ as defined after eq. (\ref{123}). We also use the
abbreviations $\bar{G}_{12}\equiv G_{12}[A-B]$,
$\bar{G}_j\equiv [A-B]^{-1}G_j$, and $\nabla ^2$ acts on either
one of the two formal arguments of $\bar{G}_j$; $\tilde{\Sigma}_1$
was defined in
(\ref{114}), $\tilde{\Sigma}_2$ follows by ``$1\leftrightarrow 2$'',
and $A,B$ are given in eqs. (\ref{106}). We remark that due to the
assumption of {\it translation invariance} of the two-point functions
their position in a product is irrelevant, if they are connected by
integrations over intermediate
coordinates.\footnote{Implementing
translation invariance into the effective action does not
interfere with taking one further functional derivative as in
deriving equations of motion, if proper care is taken of the ``zero
mode'', see the Appendix.} Furthermore, we employed eqs. (\ref{200})
from the Appendix and introduced the {\it spatial average} of a
function $f$, $\langle f\rangle\equiv\int d^dx\;f(x)/\mbox{V}_d$, with
$\mbox{V}_d\equiv\int d^dx$ denoting the volume of the
region of integration. Of course, all one- and
two-point functions in eq. (\ref{124}) are time-dependent, which we
suppressed.

We conclude this section by deriving the corresponding TDHF
equations of motion for the
{\it ten variational parameter functions} entering the effective
action in eq. (\ref{124}). In particular, these equations describe the
time-evolution of our
{\it parton model field theory} defined by eqs. (\ref{83},\ref{84})
together with eqs. (\ref{84.1}):
\beqar
\frac{\delta\Gamma}{\delta\bar{\pi}_j(x,t)}&=&0\;\;
\Longrightarrow\;\;
\dot{\bar{\phi}}_j(x,t)\; =\;\bar{\pi}_j(x,t)
\;\;. \label{125} \\ [1ex]
\frac{\delta\Gamma}{\delta\Sigma_j(x,y,t)}&=&0\;\;
\Longrightarrow\;\;
\dot{\bar{G}}_j(x,y,t)\; =\;4\tilde{\Sigma}_j\bar{G}_j|_{(x,y;t)}
\;\;. \label{126} \\ [1ex]
\frac{\delta\Gamma}{\delta\Sigma_{12}(x,y,t)}&=&0\;\;
\Longrightarrow\;\;
\partial_t\left (\bar{G}_1\bar{G}_2\bar{G}_{12}\right )_{(x,y;t)}\; =\;
2(\tilde{\Sigma}_1+\tilde{\Sigma}_2)
\bar{G}_1\bar{G}_2\bar{G}_{12}|_{(x,y;t)}
\nonumber \\
&\;&\;\;\;\;\;\;\;\;\;\;\;\;\;\;\;\;\;\;\;\;\;\;\;\;\;\;\;\;\;\;
\;\;\;\;\;\;\;\;\;\;\;\;\;\;\;\;\;\;\;
-\frac{1}{2}(G_1+G_2)\Sigma
_{12}|_{(x,y;t)}
\;\;. \label{127} \\ [1ex]
\frac{\delta\Gamma}{\delta\bar{\phi}_j(x,t)}&=&0\;\;
\Longrightarrow\;\;
\dot{\bar{\pi}}_j(x,t)\; =\;
\ddot{\bar{\phi}}_j(x,t)\; =\;
\nabla^2\bar{\phi}_j(x,t)-v_j^{(1)}-V_j^{(1)}
\label{128} \\
&\;&\;\;\;\;\;\;\;\;\;\;\;\;\;
-\frac{\hbar}{2}\{ v_j^{(3)}+V_j^{(3)}
\}\bar{G}_j(x,x,t)+\hbar\frac{d}{d\bar{\phi}_j}V^{(1,1)}
\bar{G}_1\bar{G}_2\bar{G}_{12}|_{(x,x;t)}
\;, \nonumber \eeqar{128.1}
where we used eq. (\ref{125}) to indicate that $\bar{\pi}_j$ can be
easily eliminated from the set of equations. Apart from eq. (\ref{127})
and the last term in eq. (\ref{128}),
the above equations still have a structure similar to the ones which
can be obtained from eq. (\ref{97}) in the case of a single scalar
field \cite{CM,CPS}. Note that eq. (\ref{127}) can be explicitly
written as an equation for $\dot{\bar{G}}_{12}$, eliminating
$\dot{\bar{G}}_j$ by eq. (\ref{126}).

We remark that the equations following by variation w.r.t. the
two-point functions $\Sigma_j$ and $\Sigma_{12}$, in particular,
could be derived by directly applying the respective functional
derivatives on the effective action, since {\it all} contributions
arise in this case from trace terms in eq. (\ref{124}). For the other
two-point functions, $G_j$ and $G_{12}$, there are terms with potential
insertions. Here one has to be more careful, taking into
account that the two-point functions actually depend only on {\it one}
variable due to translation invariance. The relevant
formulae are given in the Appendix.

Then, we obtain another pair of equations which generalize
the corresponding one for a single scalar field ($j'\neq j,\;j=1,2$):
\beqar
&\;&\frac{\delta\Gamma}{\delta G_j(x,y,t)}\; =\; 0\;\;
\Longrightarrow\;\;
\nonumber \\ [2ex]
&\;&\dot{\Sigma}_j(x,y,t)
\; =\; [-2\tilde{\Sigma}_j\tilde{\Sigma}_j
+\frac{1}{8}\bar{G}_j^{-2}
+\frac{1}{2}G_{j'}G_{12}\dot{\Sigma}_{12}
+G_{j'}\tilde{\Sigma}_{j'}\bar{G}_{12}\Sigma_{12}]_{(x,y;t)}
\nonumber \\ [1ex]
&\;&\;\;\;
+\left [\frac{1}{2}\nabla ^2-\frac{1}{2}\langle\{ v_j^{(2)}+V_j^
{(2)}\}\rangle\right.-
\frac{\hbar}{\mbox{V}_d}\left.\frac{1}{4}\{ v_j^{(4)}+V_j^{(4)}\}
\;\mbox{tr}\;\bar{G} _j\right ]_{(t)}\cdot\delta (x-y)
\nonumber \\
&\;&\;\;\;
+\;\langle V^{(1,1)}\rangle G_{j'}G_{12}|_{(x,y;t)}
-\frac{\hbar}{\mbox{V}_d}\left.\frac{1}{4}V^{(2,2)}\right [
4\;\mbox{tr}\; [\bar{G}_1\bar{G}_2\bar{G}_{12}]\cdot G_{j'}G_{12}
\nonumber \\
&\;&\;\;\;\;\;\;\;\;
+\;\mbox{tr}\; G_{j'}\cdot [A-B]^{-1}
+\frac{1}{2}\;\mbox{tr}\; G_j\cdot G_{j'}^{\;2}G_{12}^{\;2}F
+\;\mbox{tr}\; [\bar{G}_{j'}[A-B]^{-1}]\cdot G_1G_2\bar{G}_{12}^{\;2}
\nonumber \\
&\;&\;\;\;\;\;\;\;\;
+\;\mbox{tr}\; [G_j^{\;2}G_{j'}G_{12}^{\;2}]\cdot G_{j'}^{\;2}G_{12}^
{\;2}[A-B]^{-1}
+\frac{1}{2}\;\mbox{tr}\; [\bar{G}_j[A-B]^{-1}]\cdot G_{j'}^{\;2}
\bar{G}_{12}^{\;2}
\nonumber \\
&\;&\;\;\;\;\;\;\;\;
+4\;\mbox{tr}\; G_{12}\cdot\bar{G}_jG_{j'}^{\;2}G_{12}
+2\;\mbox{tr}\; [\bar{G}_1\bar{G}_2G_{12}]\cdot G_{j'}G_{12}[A-B]^2
\nonumber \\
&\;&\;\;\;\;\;\;\;\;
+2\;\mbox{tr}\; [G_1G_2G_{12}]\cdot G_{j'}G_{12}F
+\frac{1}{2}\;\mbox{tr}\; [G_j\bar{G}_{j'}^{\;2}G_{12}^{\;2}]\cdot
[A-B]^2
\nonumber \\
&\;&\;\;\;\;\;\;\;\;+\left.
\frac{1}{2}\;\mbox{tr}\; [G_jG_{j'}^{\;2}G_{12}^{\;2}]\cdot F
+\;\mbox{tr}\;\bar{G}_{j'}\cdot (\hat{1}+[A-B]^2)
\;\right ]_{(x,y;t)}
\;\;, \eeqar{129}
where we used formulae from the Appendix and inserted the abbreviations
introduced after eq. (\ref{124}) and $F\equiv [1+G_1G_2G_{12}^{\;2}]/
[1-G_1G_2G_{12}^{\;2}]$
to write the result as compact as possible. For the same reason we
collected the arguments from individual terms
and wrote $\hat{1}$ for a $\delta$-function in the last line.
At last, we find:
\beqar
&\;&\frac{\delta\Gamma}{\delta G_{12}(x,y,t)}\; =\; 0\;\;
\Longrightarrow\;\;
\nonumber \\ [2ex]
&\;&\dot{\Sigma}_{12}
[\hat{1}+G_1G_2G_{12}^{\;2}]|_{(x,y;t)}\; =\;
4\sum_{j=1,2}[\; -\frac{1}{2}
\tilde{\Sigma}_j\Sigma_{12}[A-B]+(\dot{\Sigma}_j
+2\tilde{\Sigma}_j^{\;2}
)G_jG_{12}
\nonumber \\
&\;&\;\;\;\;
-\left [\frac{1}{2}\nabla ^2-\frac{1}{2}\langle\{ v_j^{(2)}+V_j^
{(2)}\}\rangle\right.-
\frac{\hbar}{\mbox{V}_d}\left.\frac{1}{4}\{ v_j^{(4)}+V_j^{(4)}\}
\;\mbox{tr}\;\bar{G} _j\right ]_{(t)}\cdot G_jG_{12}\; ]_{(x,y;t)}
\nonumber \\
&\;&\;\;\;\;
-\; 2\langle V^{(1,1)}\rangle
[\hat{1}+G_1G_2G_{12}^{\;2}]|_{(x,y;t)}
+\frac{\hbar}{\mbox{V}_d}\left.V^{(2,2)}\right [
\; 2\;\mbox{tr}\; [\bar{G}_1\bar{G}_2\bar{G}_{12}]\cdot
[\hat{1}+G_1G_2G_{12}^{\;2}]
\nonumber \\ [1ex]
&\;&\;\;\;\;\;\;\;\;\;\;
+\;\mbox{tr}\; [\bar{G}_1^{\;2}\bar{G}_2^{\;2}\bar{G}_{12}[A-B]]\cdot
\bar{G}_1^{-1}\bar{G}_2^{-1}
+\;\mbox{tr}\; [\bar{G}_1\bar{G}_2G_{12}]\cdot [A-B]^2
\nonumber \\ [1ex]
&\;&\;\;\;\;\;\;\;\;\;\;
+\;\mbox{tr}\; G_{12}\cdot
G_1G_2(4[A-B]^{-1}-3)
+\;\mbox{tr}\; [G_1G_2G_{12}]\cdot
(4[A-B]^{-1}-3)
\nonumber \\
&\;&\;\;\;\;\;\;\;\;\;\;
+\{\;\mbox{tr}\;\bar{G}_1\cdot
G_2G_{12}+
\;\mbox{tr}\;G_1\cdot
G_2G_{12}(\hat{1}+\frac{1}{2}F)+
\;\mbox{tr}\; [G_1G_2^{\;2}G_{12}^{\;2}]\cdot\bar{G_1}G_{12}
\nonumber \\
&\;&\;\;\;\;\;\;\;\;\;\;\;\;\;+\frac{1}{2}
\left.\mbox{tr}\; [\bar{G}_1[A-B]^{-1}]\cdot G_2\bar{G}_{12}[A-B]
\;\}\; +\;\{\; ``1\leftrightarrow 2\mbox{''}\;\}\;\right ]_{(x,y;t)}
\;\;, \eeqar{130}
where the same comments as after eq. (\ref{129}) apply; in particular,
note that $\nabla ^2$ acts on either one argument of the following
term, as before.

Thus, with eqs. (\ref{125}) - (\ref{130}) we obtained a closed set
of {\it ten coupled non-linear equations} which describe the
time-evolution of any theory of {\it two coupled scalar fields} in
terms of one- and two-point Wightman functions (TDHF approximation).
They are local in time and of first order in time
derivatives (see, however, eq. (\ref{128}) for the elimination of
$\bar{\pi}_j$ and consequently $\partial _t\bar{\phi}_j\rightarrow
\partial _t^{\;2}\bar{\phi}_j$).

Generally, one has to expect the usual
ultraviolet divergences in terms like
$\mbox{tr}\;G(t)=\int d^dx\; G(x,x)$ corresponding to expectation values
of two fields at coinciding points. We assume them to be regulated by a
high-momentum cut-off at present. Note that by their definition via
the Momentum Space Mode Separation in Sec. 4.1 the gluonic environment
modes are automatically regulated in this respect. Therefore, one
could essentially take over the separate and only necessary
{\it renormalization of the scalar parton} sector from Refs.
\cite{CM,CPS}. We plan to come back to this feature in the sequel to
the present work. There we also have to specify physical {\it initial
conditions} for the set of parton-environment TDHF equations and
study the restrictions imposed by renormalization.

In passing we remark that due to translation invariance eqs. (\ref{125})
- (\ref{130}) simplify considerably after a {\it Fourier transformation}.
 From multiple convolution type spatial integrals involving two-point
functions there is always only one final momentum integration left over,
 cf. e.g. eqs. (\ref{107},\ref{111}) where this was employed before.
Incidentally, for any translation invariant two-point $F$ function
here, with
\beq
F(p;t)\;\equiv\;\int d^dx\;\mbox{e}^{-ip\cdot y}\; F(y;t)\; =\;
\int d^dx\;\mbox{e}^{-ip\cdot y}\; F(x+\half y,x-\half y;t)
\;\;, \eeq{131}
the Fourier transform coincides with the non-covariant {\it Wigner
function}. Therefore, after a straightforward Fourier
transformation, our set of equations at the same time presents the
{\it transport theory} \cite{us} for any two coupled scalar fields in
TDHF approximation. It is obtained here in the non-covariant form
suitable for initial value problems and should allow interesting
comparisons with other transport equations derived in weak-coupling
perturbation theory.

Finally, we point out two special cases of the
parton-environment TDHF equations, which are of interest in their
own right:
\vskip .15cm
\noindent
{\bf (a)} Setting $\bar{\phi}_j\equiv 0\equiv\bar{\pi}_j$, i.e.
{\it no classical mean fields}, the set of equations reduces to
{\it six} coupled equations. This may be relevant for a short-lived
system with no such fields in the initial state which is in some
sense dominated by incoherent radiation, i.e., if there is no time to
develop classical expectation values of the fields. It might also
apply to a densely (in phase space) populated parton system, if
(and only {\it IF}) the effect of such a ``quasi high-temperature''
state is to ``melt'' the classical fields.
\vskip .15cm
\noindent
{\bf (b)} Conversely, if we constrain one part of the complex system
to follow the classical equations of motion, i.e. to be completely
{\it mean field dominated}, by setting e.g. $G_2\equiv 0\equiv\Sigma_2$,
then the set of equations again reduces to only {\it six} coupled
equations, since the dependence on the correlation functions
$G_{12}$ and $\Sigma_{12}$ is automatically eliminated from the
effective action, eq. (\ref{124}). This is in agreement with our
conclusions concerning {\it entropy production}, cf. eqs.
(\ref{107},\ref{111}) and point {\bf III.} in Sec. 4.3, which is
identically {\it zero} here. Obviously, this special case (i.e. the
QCD analogue to be worked out) cannot be
relevant for high-multiplicity events in strong interactions at high
energy and the parton model as applied to such events, in particular.
\vskip .15cm
\noindent
We further discuss these issues in the following section.

\yoursection{5. Conclusions}

At the outset we apologize to all practitioners and followers of the
QCD parton model. We did not consider QCD partons and their gluonic
environment yet. However, our considerations and the basic idea
towards a solution of the about 40 years old ``{\it entropy puzzle}''
in strong interactions at high energy are motivated by what we believe
to be important features of QCD. This explains our abuse of the words
``parton'' and ``gluonic'' environment modes.

To begin with, throughout this work and in Sec. 2, in particular, we
consider the von Neumann or {\it statistical entropy} defined in
terms of the relevant density matrix, see e.g. eq. (\ref{17}), the
part of Sec. 3 following point {\bf (iii)}, and especially Sec. 4.3.
For technical reasons we often employ the {\it linear entropy},
eq. (\ref{18}), which we show to provide a lower bound on the
statistical entropy, see eq. (\ref{110}). Concerning {\it quantum}
properties of the system to be characterized by the entropy, both
definitions are equally valid for our purposes. In particular, both
measure the {\it impurity} of the parton density matrix.

The core of the entropy problem is to try and understand how (an
idealized example) two hadronic scattering in-states undergoing a
hard interaction, i.e. a quantum mechanically {\it pure initial
state}, can result in a high-multiplicity event corresponding to a
highly impure (more or less thermal) density matrix on the parton
level before hadronization.

Thus, attempts to associate the apparent entropy production with a
``coarse graining'' either in the phase space of the observed system
or due to the finite resolution in any experimental measurement,
which both may be useful to characterize derived ``macroscopic''
aspects of such reactions, seem to miss the point. There one gives up
from the beginning
the possibility to understand on a fundamental dynamical level how a
{\it complex pure-state quantum system} can produce classical behaviour,
 i.e. an impure density (sub-)matrix with {\it decoherence} of
associated (parton) wave functions and {\it entropy} production.

In Sec. 2 we introduce a convenient general framework in terms of
{\it Schmidt} and {\it pointer states} to analyze the consequences of
a dynamical separation of a complex (possibly strongly interacting)
system into an observable (``partonic'') subsystem plus unobservable
(``gluonic'') environment modes \cite{Zu0,Om,GH,Alb}. Entropy production
 due to {\it environment-induced quantum decoherence} in the observable
subsystem arises naturally, {\it IF} such a separation of the total
``closed'' system into an ``open'' subsystem and its environment is
{\it dynamically} realized. There is no guarantee for this to happen
and the existence of the associated (almost) classically behaving
(partonic) pointer states, i.e. states the quantum superpositions of
which are dynamically suppressed, has to be viewed as a particular
feature of the system.

Such a major miracle seems to be installed in
QCD \cite{AMuller}: for example, the fact that deep-inelastic
scattering can be described accurately in terms of a hard scattering
cross section and structure functions. The latter are {\it decoherent}
one-particle {\it probability distributions}. Therefore, some ``secret
agent'' has to effect this extremely efficient decoherence process,
once a parton is knocked out of its coherent initial state wave
function. Our point of view here does not conflict with the standard
parton picture, in particular, the applicability of plane-wave states
representing partons entering or leaving Feynman diagrams.\footnote{I
thank R. Baier for insisting with his questions about this.} Any basis
should do. However, the miracle consists in the fact that typically the
initial state of a hadronic scattering reaction can be described by
structure functions which correspond to {\it diagonal} density
matrices, i.e. without interference terms (which {\it always} can be
achieved formally by the Schmidt decomposition procedure, however,
only at {\it one instant of time}, cf. Sec. 2).

In terms of the split hamiltonian in eq. (\ref{20}), the bound
partonic initial state wave function diagonalizes $\hat{\mbox{H}}_
{\cal P}+\hat{\mbox{H}}_{\cal PG}$. Once it is perturbed by an
external (e.g. electromagnetic) interaction, the interaction with
the gluonic environment starts to ``rattle and shake'' the perturbed
wave function so strongly that it decays very fast into a decoherent
superposition of states as reflected in a {\it structure
function}.

We conjecture that gluonic modes (neglecting quarks for now) which
are close to {\it constant field configurations} form an essentially
unobservable sector of the total Hilbert space of a QCD system. The
associated Momentum Space Mode Separation is discussed in more detail
in Sec. 4.1. They act as an active environment on the
observable\footnote{In the sense of parton-hadron duality or
deep-inelastic scattering.} partonic sector, to which they are strongly
coupled via low-$|Q|^2$ processes. Presently we explore
this picture in two simple
models.

In Sec. 3 we study a non-relativistic parton coupled to a gluonic
environment in close analogy to an electron coupled to the quantized
electromagnetic field \cite{BC,HA}. However, we deliberately change
the spectral density of the environmental oscillators to be dominated
in the infrared. We calculate the time-dependent parton density
matrix by integrating out the environmental degrees of freedom
exactly with the Feynman-Vernon influence functional technique
\cite{Feyn,Grab}. In a {\it short-time strong-coupling} limit, which
has not been of interest in studies of quantum Brownian motion so far
\cite{Grab}, we find the following results analytically (cf. fig. 1
and points {\bf (i)} - {\bf (iii)} in Sec. 3.2):
\vskip .15cm
\noindent
Gaussian parton wave packets experience {\it friction} and {\it
localization} and their coherent {\it superpositions decohere}.
\vskip .15cm
\noindent
All effects are governed by a non-perturbative time-scale $f_+^{-1}
\ll 1$fm/c, eq. (\ref{64}). Thus, a parton following the classical
trajectory with the center of its wave packet is slowly decelerated,
which can be interpreted as being due to the emission of infrared
gluons, i.e. the excitation of environment modes in this model. More
surprising is the localization effect, i.e., the quantum mechanical
spreading of the wave packet can be suppressed or even reversed
depending on the choice of parameters. The initial wave function
obtains an almost soliton like character through the strong interaction
with the environment. Finally, considering the superposition of
wave packets, we find that their decoherence actually happens in the
short-time strong-coupling limit. Consequently, in accordance with
the general considerations of Sec. 2, we ``see'' and can calculate
how {\it entropy} is produced in this toy model on a short time scale.

All the above effects, if recovered in QCD eventually, seem to be
very important e.g. for a justification of the classical cascade
approach to multiple parton scattering \cite{GM}. Then, a parton
propagating in space-time behaves essentially like a {\it classical
particle} in between perturbative hard scatterings.

In Sec. 4 we set up a fairly general model of two coupled scalar
fields representing partons and gluonic environment modes, respectively
(see Sec. 4.1 for details). In this case, due to the non-linear
interactions, the environment degrees of freedom cannot be integrated
out as before. Therefore, our main concern here is to formulate a
tractable approach to the time-evolution problem in this quantum
field theory, which is not restricted to perturbatively small couplings,
 in particular. We employ Dirac's variational principle
\cite{Dirac,Jackiw}, see Sec. 4.2, and describe the complex system
in terms of its {\it ten one- and two-point Wightman functions}
(TDHF approximation). This corresponds to the most general gaussian
wave functional, eq. (\ref{100}), in the field theory Schr\"odinger
picture. In Sec. 4.4 we derive the related Cornwall-Jackiw-Tomboulis
type {\it effective action} \cite{Corn}, eq. (\ref{124}), governing
the dynamics and obtain the equations of motion from it.

An attractive feature of having an explicit wave functional in hand
is that it allows to calculate the {\it partonic density functional}
here quite generally, eqs. (\ref{102}) - (\ref{104}) or eqs. (\ref{112})
 - (\ref{114}), without having to solve the complicated dynamics first
of all. Thus, we obtain in Sec. 4.3 a {\it model-independent} result
for the {\it entropy} in terms of two-point functions, eqs.
(\ref{107},\ref{111}), which causes several interesting observations,
see points {\bf I.} - {\bf IV.} there, which are presumably of a
general nature:
\vskip .15cm
\noindent
{\bf 1.} If there are {\it no quantum correlations} between the
partons and the gluonic environment, then there is {\it no parton
decoherence} and {\it no entropy production}. Again, this is in
accordance with the general considerations in Sec. 2.
\vskip .15cm
\noindent
{\bf 2.} If one assumes a strictly {\it classical gluonic environment},
i.e. following the classical equations of motion, then there is {\it no
parton decoherence} and {\it no entropy production}.
\vskip .15cm
This latter observation is in interesting contrast to recent attempts
to explain entropy production and thermalization in QCD as being
due to the non-perturbative chaotic dynamics of purely classical
Yang-Mills fields, see Ref. \cite{MT} and references therein.
There the above mentioned ``coarse
graining'' of the classical phase space covered by an {\it ensemble}
of identical systems\footnote{Of course, this does {\it not} represent
a {\it pure} quantum state and, thus, the resulting ``thermalization''
seems to be a very plausible consequence of the mixing property of a
strongly chaotic system, i.e. the filamentation of the phase space
occupied by the ensemble.} is necessary to deduce a {\it classical
entropy}. The results are very suggestive as compared to high-temperature
 field theory. However, in both cases decoherence is put in by hand.
Their relevance for the von Neumann entropy considered here
and the most important {\it dynamical decoherence} phenomena in a
{\it quantum} parton system, in particular, seems to be unclear
at present.

Our results in Sec. 4 will be the starting point of the sequel to
this work, where we plan to study physical initial conditions for
the parton-environment system and their
relation to model structure functions. Then, of course, the
time-evolution of the system following the equations of motion,
eqs. (\ref{125}) - (\ref{130}), will be most interesting to consider.
Any solutions will provide the explicit time-dependence
of entropy production
according to eqs. (\ref{107},\ref{111}), especially for early stages
of a reaction.

Let us summarize our present point of view by saying
that a parton appears like a parton only because it feels the
gluonic environment which manifests its {\it strong} non-perturbative
interactions caused by the running coupling
on a short time scale ($\ll 1$fm/c) through the
induced decoherence properties of partons and the associated
observable entropy production.

Finally, it seems worthwhile to mention a few other potential
applications of
our results from Sec. 4.
They provide a first step to study the analogue of
{\it quantum Brownian motion} \cite{Grab} in the context of field theory.
In particular, one may study the time-dependence of a Higgs model
type phase transition under the influence of the interaction with
a perturbing environment field. The possibility of such a
{\it non-equilibrium phase transition} is contained already in our
simplest
specification of the interactions in eqs. (\ref{84.1}). It has been
another long-standing problem to study changes of signals, which are
thought to characterize an equilibrium phase transition, when they come
 from an out-of-equilibrium system such as the early stage of a
high-energy nuclear collision. Recently, following the suggestion
of the possible formation of a disordered chiral condensate
in relativistic heavy-ion collisions, it has been realized
that the condensate field actually presents an {\it open system}, see
Refs. \cite{AK} and numerous references therein, to which our
methods can be applied.

\vskip 0.2cm
\noindent
{\bf Acknowledgements}

I thank the organizers of the
Workshop on Pre-equilibrium Parton Dynamics in Heavy-Ion Collisions
at LBL (Berkeley), M. Gyulassy, B. M\"uller and X.-N. Wang,
for their support and providing the enjoyable atmosphere, where
this work originated.
I thank K. Geiger for explaining me the machinery of the
parton model and numerous enthusiastic conversations about and
beyond this subject, as well as for many helpful suggestions improving
the manuscript.
The kind hospitality of the members of the theory group at the
Universit\'e Paris XI (Orsay) and the
opportunity to present this work to a very stimulating audience there
are much appreciated.

\yoursection{Appendix}

We list several formulae which are useful for the calculations in
Sec. 4.4.\footnote{C. Wetterich helped with a remark here at the
right moment.}
They are valid for any translation invariant symmetric
two-point functions $X,Y,Z$, and any integrable
function $\cal V$ (in any dimension):
\beqar
I&\equiv&\int dx\; {\cal V}(x)\cdot XYZ|_{(x,x)}
\nonumber \\
&=&\int dx\; dx_1dx_2\; {\cal V}(x)X(x-x_1)Y(x_1-x_2)Z(x_2-x)
\nonumber \\
&=&\int dx\; {\cal V}(x)\cdot XZY|_{(x,x)}
\;\;\;
\left [\;\mbox{or any other permutation of}\; XYZ\;\right ]
\nonumber \\
&=&\int dx\; {\cal V}(x)\;\mbox{tr}\; [XYZ]/vol
\;\;, \eeqar{200}
where $vol\equiv\int dy$ denotes the volume of the region of
integration. Then, functional derivatives w.r.t. one of the two-point
functions can be calculated in two ways employing the {\it trace formula}
 in (\ref{200}):
\beqar
\frac{\delta}{\delta Z(z)}\; I&=&
\int dx'\; {\cal V}(x')\;
\frac{\delta}{\delta Z(z)}
\;\mbox{tr}\; [XYZ]/vol
\nonumber \\
&=&
\int dx'\; {\cal V}(x')\; XY|_{(x,y)}\;\;\;\;
\left [\;\mbox{for}\;\; z\equiv x-y\;\right ]
\nonumber \\
&=&
\int dx'\; {\cal V}(x')\;
\frac{\delta}{\delta Z(x,y)}
\;\mbox{tr}\; [XYZ]
\;\;. \eeqar{201}
Dividing out the constant factor, one obtains:
\beq
\frac{\delta}{\delta Z(x-y)}\;\mbox{tr}\; [XYZ]\; =\;
vol\cdot\frac{\delta}{\delta Z(x,y)}\;\mbox{tr}\; [XYZ]
\;\;, \eeq{202}
which clearly exhibits the ``zero mode'' factor connecting the
functional derivative which takes translational invariance into
account with the one which does not.

Finally, with the definitions of $A,B$ and $\bar{G}_j,\bar{G}_{12}$
given in eqs. (\ref{106}) and after eq. (\ref{124}), respectively,
one finds the following simple results:
\beq
\mbox{tr}\; [X
\frac{\delta \bar{G}_j}{\delta G_j(x,y)}]\; =\;\frac{X}{[A-B]^2}|_{(x,y)}
\;\;,\;\;\;
\mbox{tr}\; [X
\frac{\delta \bar{G}_{12}}{\delta G_{12}(x,y)}]\; =\;
X(3[A-B]-2)|_{(x,y)}
\;\;, \eeq{203}
where the permutability of translation invariant symmetric two-point
functions in a product, e.g. $XYZ|_{(x,y)}=XZY|_{(x,y)}=\ldots\;$
(cf. also eq. (\ref{200})), is crucial. Similar useful relations are
easily obtained for any power of $[A-B]$ multiplying $G_j$ or $G_{12}$.

\vskip 2.0cm
\noindent
{\bf Figure captions:}
\vskip 0.15cm
\noindent
Fig. 1: Deceleration and localization of a non-relativistic parton in
a gluonic environment. Shown are the unique deceleration $v/v_0$,
eq. (\ref{66}),
and the relative change of the width $(w-\beta)/\beta$, eq. (\ref{68.1}),
 of the wave packet as a function of $t_+\equiv f_+t$.
We set $\alpha _<=0.1\alpha$; see eqs. (\ref{68.2}) for
the definitions of the parameters.
For comparison free particle results (dashed curves) are given.
\vskip 0.15cm
\noindent
Fig. 2: Qualitative picture of Momentum Space Mode Separation:
``Unobservable'' gluonic environment modes are confined to the inner
region II (roughly $k_0^{\;2}<m_\pi^{\;2}$ and $\vec{k}^2<\Lambda_{QCD}
^{\;2}$). Partons live in the outer region I.
\eject

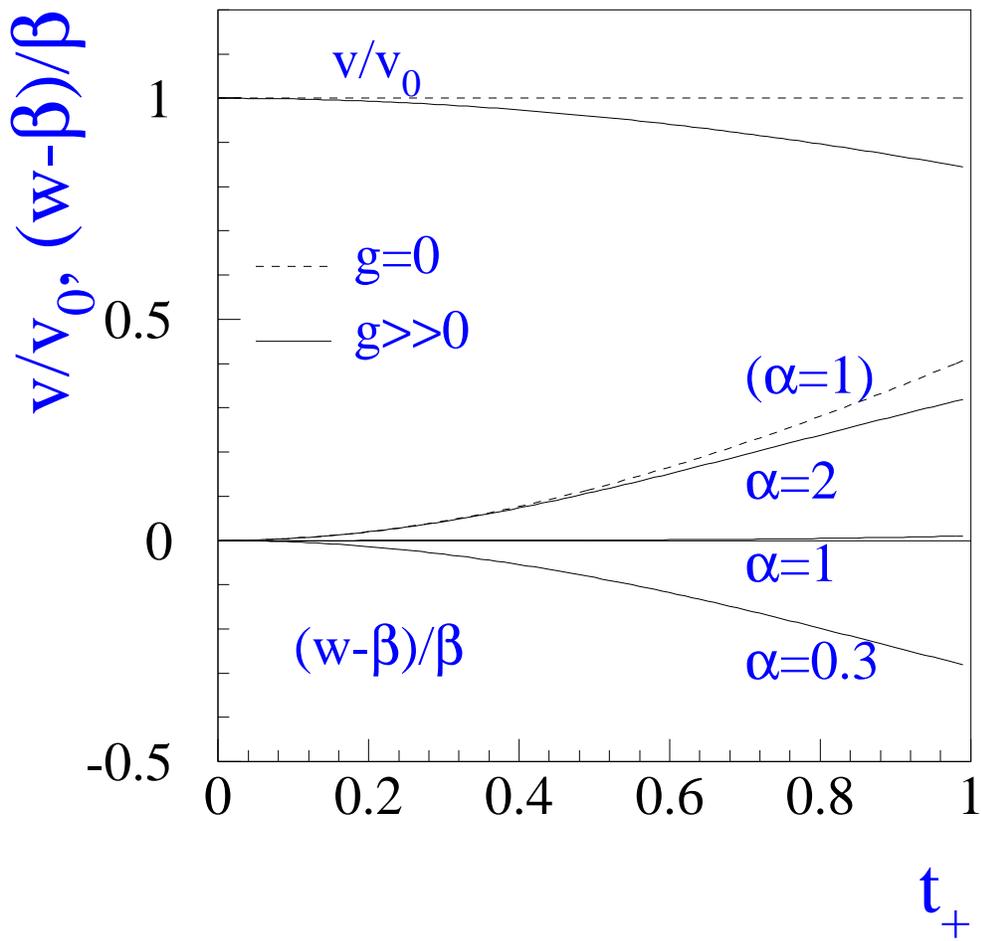
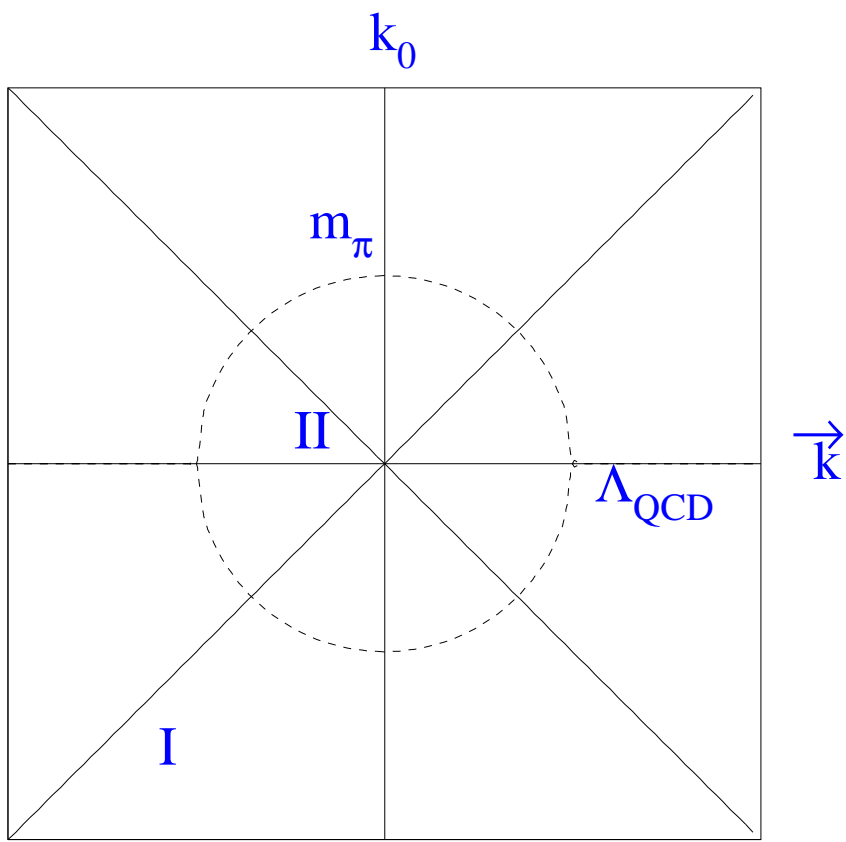
\end{document}